\documentclass[10pt,journal,onecolumn,twoside]{IEEEtran}
\newcommand{\B}[1]{\boldsymbol{#1}}

\usepackage{amsmath}
\usepackage{amssymb}
\usepackage{epsfig}
\usepackage{psfrag}

\begin{document}
\title{Low SNR Asymptotic Rates of Vector Channels with One-Bit Outputs}
\author{Amine~Mezghani,~\IEEEmembership{Member,~IEEE,} 
        Josef~A.~Nossek,~\IEEEmembership{Life Fellow,~IEEE,} A.~Lee~Swindelhurst~\IEEEmembership{Fellow,~IEEE} 
\thanks{A. Mezghani was with the University of California, Irvine, USA, and is now with the University of Texas at Austin, USA. J.~A.~Nossek is with the Technical University of Munich Germany, and the Federal University of Cear\'a, Fortaleza, Brazil.  A. L. Swindelhurst is with the Department of EECS, University of California, Irvine, USA, and is also a Hans Fischer Senior Fellow of the Institute for Advanced Study at the Technical University of Munich. E-mail: amine.mezghani@tum.de.}
\thanks{This work was supported by the National Science Foundation (NSF) under ECCS-1547155 and by the German Research Foundation (DFG) under the priority program SPP1202.}
\thanks{Parts of this work was published in \cite{mezghaniisit2007,mezghaniisit2008,mezghaniisit2009}, where some proofs were missing, and in particular, the main theorem in Section~\ref{entropy_theorem} was not stated and the application to massive MIMO was not investigated.}
\thanks{Manuscript received December xx, xxxx; revised January xx, xxxx.}}
\markboth{SUBMITTED TO THE IEEE TRANSACTIONS ON INFORMATION THEORY, December~2017}{Mezghani, Nossek \MakeLowercase{\textit{and}} Swindlehurst: Low SNR Asymptotic Rates of Vector Channels with single Bit Outputs}
\maketitle 
\begin{abstract}
We analyze  the performance of multiple-input multiple-output (MIMO) links with one-bit output quantization in terms of achievable rates and characterize their performance loss compared to unquantized	systems for general channel statistical models and general channel state information (CSI) at the receiver. One-bit ADCs are particularly suitable for large-scale millimeter wave MIMO Communications (massive MIMO) to reduce the hardware complexity. In such applications,  the signal-to-noise ratio per antenna is rather low due to the propagation loss. Thus, it is crucial to analyze the performance of MIMO systems in this regime by means of information theoretical methods.    Since an exact and general information-theoretic analysis is not possible, we resort to the derivation of a general  asymptotic expression for the mutual information in terms of a second order expansion around zero SNR. We show that up to second order in the SNR, the mutual information of a system with two-level (sign) output signals incorporates only a power penalty factor of $\pi/2$ (1.96 dB) compared to system with infinite resolution for all channels of practical interest with perfect or statistical CSI. 
An essential aspect of the derivation is that we do not rely on the common pseudo-quantization noise model.
\end{abstract}
\begin{IEEEkeywords}
Massive MIMO communication, Broadband regime, One-bit Quantization, Mutual information, Optimal input distribution, Ergodic capacity, Millimeter-wave Communications.
\end{IEEEkeywords}
\section{Introduction}
In this paper, we investigate the theoretically achievable rates under one-bit analog-to-digital conversion (ADC) at the receiver for a wide class of channel models. To this end, we consider general multi-antenna communication channels with coarsely quantized outputs and general communication scenarios, e.g. correlated fading, full and statistical channel state information (CSI) at the transmitter and the receiver, etc.. Since  exact capacity formulas are intractable in such quantized channels, we resort to a low signal-to-noise ratio (SNR) approximation and lower bounds on the channel capacity to perform the analysis.  Such mutual information asymptotics can be utilized to evaluate the performance of quantized output channels or design  and optimize the system in practice. Additionally, the low SNR analysis under coarse quantization is useful in the context of large scale (or massive) multiple-input multiple-output (MIMO) \cite{MarzettaITA,Mar10,LuLu2014} and millimeter-wave (mmwave) communications \cite{Madhow2008,Boccardi2014,swindlehurst,Roh14,Ran14}, considered as key enablers to achieve higher data rates in future wireless networks.  In fact,  due to high antenna gains possible with massive MIMO and the significant path-loss at mmwave frequencies, such systems will likely operate at rather low SNR values at each antenna, while preferably utilizing low cost hardware and low resolution ADCs, in order to access all available dimensions even at low precision. Our asymptotic analysis demonstrates that the capacity degradation due to quantized sampling is surprisingly small in the low SNR regime for most cases of practical interest.    

\subsection{Less precision for more dimensions: The motivation for coarse quantization}

The use of low resolution (e.g., one-bit) ADCs and DACs is a potential approach to significantly reducing cost and power consumption in massive MIMO wireless transceivers. It was proposed as early as 2006 by \cite{ivrlac2006}-\nocite{boyd,boyd_2}\cite{nossek} in the context of conventional MIMO.  In the last three years however, the topic has gained significantly increased interest by the research community \cite{mezghani2007}-\nocite{mezghaniisit2007,mezghani_ICASSP2008,mezghaniisit2008,mezghaniisit2009,mezghani_itg_2010,koch,jaspreet_tcom09,singh_isit,koch2,zeitler,nakamura_isita,mezghaniisit2010,mezghani2008,mezghani_2012_isit,zeitler2,wang14,wang15,hea14,Studer_2016,Jacobsson_2017,singh2009multi,mezghani2012an,jianhua2014high,chiara2014massive,jianhua2014channel,ning2015mixed,jacobsson2015one,juncil2015near,Wen_2015,mollen2016performance_WSA,mollen2016one,mollen2016performance,Yongzhisam,YongzhiGlobecom,Yongzhiuplink,Stoeckle_2016}\cite{Rassouli} as an attractive low cost solution for large vector channels.
In the extreme case, a one-bit ADC consists of a simple comparator and consumes negligible power. One-bit ADCs do not require an automatic gain control and the  complexity and power consumption of the gain stages required prior to them are substantially reduced \cite{donnell}. Ultimately, one-bit conversion is, in view of current CMOS technology, the only conceivable option for a direct mmwave bandpass sampling implementation close to the antenna, eliminating the need for power intensive radio-frequency (RF) components such as mixers and oscillators. In addition, the use of one-bit ADCs not only simplifies the interface to the antennas by relaxing the RF requirements but also simplifies the interface between the converters and the digital processing unit (DSP/FPGA). For instance, the use of 10-bit converters running at 1 Gsps for 100 antennas would require a data pipeline of 1 Tbit/s to the FPGAs and a very complex and power consuming interconnect.  By using only one-bit quantization the speed and complexity are reduced by a factor of 10. Sampling with one-bit ADCs  might therefore qualify as a fundamental research area in communication theory.

\begin{figure}[htb]
\centering
\psfrag{Spectral Efficiency}[c][c]{Spectral Efficiency $C$}
\psfrag{Eb/N0}[c][c]{Energy Efficiency $E_b/N_0={\rm SNR}/C$ in dB}
\centerline{\includegraphics[width=3.4in]{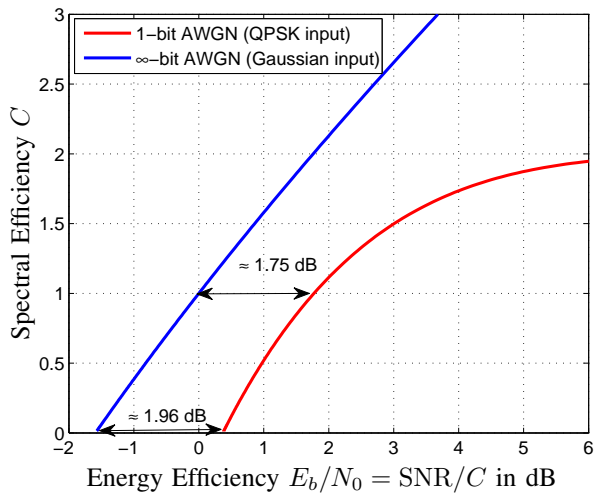}}
\caption{Spectral efficiency versus energy efficiency for One- and Infinite-Bit Quantization in AWGN channels.}
\label{SE_vs_EE}
\end{figure}

Even though use of only a single quantization bit, {\em i.e.}, simply the sign of the sampled signal, is a severe nonlinearity, initial research has shown that the theoretical ``best-case'' performance loss that results with a one-bit quantizer is not as significant as might be expected, at least at low SNRs where mmwave massive MIMO is expected to operate, prior to the beamforming gain, which can still be fully exploited. This is also very encouraging in the context of low-cost and low-power IoT devices which will also likely operate in relatively low SNR regimes. Figure~\ref{SE_vs_EE} shows how the theoretical spectral efficiency versus energy efficiency ($E_b/N_0$) of a one-bit transceiver that uses QPSK symbols in an additive white Gaussian noise (AWGN) channel compares with that of an infinite-precision ADC using a Gaussian input, i.e. the Shannon limit $\log_2(1+{\rm SNR})$. In fact, the capacity of the one-bit output AWGN channel is achieved by QPSK signals and reads as \cite{Viterbi,singh_isit}
\begin{equation}
C_{\rm 1-bit}=2\left(1- H_b(\Phi(\sqrt{ \textrm{SNR}}))  \right),
\end{equation}
where we make use of the binary entropy function $H_b(p)=-p \cdot \log_2 p - (1-p) \cdot \log_2 (1-p)$ and the cumulative Gaussian distribution $\Phi(z)$.
Surprisingly, at low SNR the loss due to one-bit quantization is approximately equal to only $\pi/2$ (1.96dB) \cite{Viterbi,verdu} and actually decreases to roughly  1.75dB at a spectral efficiency of about 1 bit per complex dimension, which corresponds to the spectral efficiency of today's 3G systems.

Even if a system is physically equipped with higher resolution converters and high performance RF-chains, situations may arise where the processing of desired signals must be performed at much lower resolution, due for instance to the presence of a strong interferer or a jammer with greater dynamic range than the signals of interest. In fact, after subtracting or zero-forcing the strong interferer, the residual effective number of bits available for the processing of other signals of interest is reduced substantially.  Since future wireless systems must operate reliably even under severe conditions in safety-critical applications such as autonomous driving, investigating communication theory and signal processing under coarse quantization of the observations is crucial.

\subsection{Related Work and Contributions}
Many contributions have studied MIMO channels operating in Rayleigh fading environments in the unquantized (infinite resolution) case,  
for both the low SNR \cite{verdu}-\nocite{prelov,rao,wu,sethuraman}\cite{durisi} and high SNR \cite{zheng2} regimes. Such asymptotic analyses are very useful since characterizing the achievable rate for the whole SNR regime is  in general intractable. This issue becomes even more difficult in the context of one-bit quantization at the receiver side, apart from very special cases. 
In the works \cite{nossek,mezghaniisit2007}, the effects of quantization were studied from an information theoretic point of view for  MIMO systems, where the channel is perfectly known at the receiver. These works demonstrated that the loss in channel capacity due to coarse quantization is surprisingly small at low to moderate SNR. In \cite{mezghaniisit2008,mezghaniisit2009}, the block fading single-input single-output (SISO) non-coherent channel was studied in detail. The work of \cite{mezghani_2012_isit} provided a general capacity lower bound for quantized MIMO and general bit resolutions that can be applied for several channel models with perfect CSI, particularly with correlated noise. The achievable capacity for the AWGN channel with output quantization has been extensively investigated in \cite{jaspreet_tcom09,singh_isit}, and the optimal input distribution was shown  to be discrete. 
The authors of \cite{koch} studied the one-bit case in the context of an AWGN channel and showed that the capacity loss can be fully recovered when using asymmetric quantizers. This is however only possible at extremely low SNR, which might not be useful in practice. In \cite{koch1}, it was shown that, as expected, oversampling can also reduce the quantization loss in the context of band limited AWGN channels. In \cite{zeitler,zeitler2}, non-regular quantizer  designs for maximizing the information rate are studied  for intersymbol-interference channels. More recently, \cite{hea14} studied bounds on the achievable rates of MIMO channels with one-bit ADCs and perfect channel state information at the transmitter and the receiver, particularly for the multiple-input single-output (MISO) channel. The recent work of \cite{Rassouli} analyzes the sum capacity of the two-user multiple access SISO AWGN channel, which turn to be achievable with time division and power control. 

Motivated by these works, we aim to study and characterize the communication performance of point-to-point MIMO channels with general assumptions about the channel state information at the receiver taking into account the  1-bit quantization as a deterministic operation. In particular, we derive asymptotics for the mutual information up to the second order in the SNR and study the impact of quantization. We show that up to second order in SNR for all channels of practical interest, the mutual information of a  system with two-level  (1-bit sign operation) output signals incorporates only a power penalty of  $\frac{\pi}{2}$ (-1.96 dB) compared to a system with infinite resolution. Alternatively, to achieve the same rate with the same power up to the second order as in the ideal case, the number of one-bit output dimensions has to be increased by a factor of $\pi/2$ for the case of perfect CSI and at least by $\pi^2/4$ for the  statistical CSI case, while essentially no increase in the number of transmit dimensions is required.  We also characterize analytically the compensation of the quantization effects by increasing the number of 1-bit receive dimensions to approach the ideal case. 
 
\par This paper is organized as follows: Section \ref{section:scmodel} describes the system model. Then,  Section~\ref{entropy_theorem} provides the main theorem consisting of a second order asymptotic approximation of the entropy of one-bit quantized vector signals. In Section~\ref{section:mutual}, we provide a general expression for the mutual information between the inputs and the quantized outputs of the MIMO system with perfect channel state information, then we expand that into a Taylor series up to the second order in the SNR. In Section \ref{section:mutual_n}, we extend these results to elaborate on the asymptotic capacity of 1-bit MIMO systems with statistical channel state information including Rayleigh flat-fading environments with delay spread and receive antenna correlation. 

\subsection{Notation}
 Vectors and matrices are denoted by lower and upper case italic bold letters.  The operators $(\bullet)^\mathrm {T}$, $(\bullet)^\mathrm {H}$, $\textrm{tr}(\bullet)$, $(\bullet)^*$, $\text{Re}(\bullet)$ and $\text{Im}(\bullet)$ stand for transpose, Hermitian (conjugate transpose), matrix trace, complex conjugate, real and imaginary parts of a complex number, respectively.  The terms $\B{0}_M$  and $\B{1}_M$ denote the $M$-dimensional vectors of all zeros and all ones, respectively, while $\B{I}_M$ represents the identity matrix of size $M$. The vector $\boldsymbol{x}_i$ is the $i$-th column of  matrix $\B{X}$ and $x_{i,j}$ denotes the ($i$th, $j$th) element, while $x_i$ is the $i$-th element of the vector $\B{x}$. The operator $\textrm{E}[\bullet]$ stands for expectation with respect to all random variables,  while the operator $\textrm{E}_{s|q}[\bullet]$ stands for the expectation with respect to the random variable $s$ given $q$. In addition, $\B{C}_x={\rm E}[\B{x}\B{x}^{\rm H}] - {\rm E}[\B{x}]{\rm E}[\B{x}^{\rm H}]$ represents the covariance matrix of $\B{x}$ and $\B{C}_{xy}$ denotes $\mathrm {E}[\B{x}\B{y}^{\rm H}]$. The functions $P(s)$ and $P(s|q)$  symbolize the joint probability mass function (pmf) and the conditional pmf of $s$ and $q$, respectively. Additionally, $\textrm{diag}(\boldsymbol{A})$ denotes a diagonal matrix containing only the diagonal elements of $\boldsymbol{A}$ and $\textrm{nondiag}(\boldsymbol{A})=\boldsymbol{A}-\textrm{diag}(\boldsymbol{A})$.  Finally, we represent element-wise multiplication and the Kronecker product of vectors and matrices by the operators "$\circ$" and "$\otimes$", respectively.  

\label{section:introduction}
\section{System Model}
\label{section:scmodel}
\begin{figure}[h]
\begin{center}
\psfrag{H}[c][c]{$\B{H}$}
\psfrag{G}[c][c]{$\B{G}$}
\psfrag{xd}[c][c]{$\B{\hat{x}}$}
\psfrag{x}[c][c]{$\B{x}$}
\psfrag{y}[c][c]{$\B{r}$}
\psfrag{r}[c][c]{$\B{y}$}
\psfrag{Q[.]}[c][c]{$Q(\bullet)$}
\psfrag{n}[c][c]{$\B{\eta}$}
\psfrag{M}[c][c]{$M$}
\psfrag{N}[c][c]{$N$}
\psfrag{SNR}[c][c]{$\sqrt{\textrm{SNR}}$}
{\epsfig {file=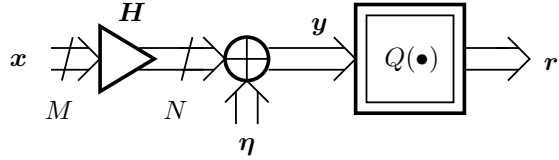, width = 7.5cm}}
\caption{Quantized MIMO System}
\label{downlink_figure}
\end{center}
\end{figure}

We consider a  point-to-point quantized MIMO channel with $M$ transmit dimensions (e.g. antennas or, more generally, spatial and temporal dimensions) and $N$  dimensions at the  receiver. Fig.~\ref{downlink_figure} shows the general form of a quantized MIMO system, where $\B{H} \in \mathbb{C}^{N\times M}$ is the  channel matrix, whose distribution is known at the receiver side. The channel realizations are in general unknown to both the transmitter and receiver, except for the ideal perfect CSI case. The vector $\B{x} \in \mathbb{C}^{M}$ comprises the $M$ transmitted symbols, assumed to be subjected to an average power constraint ${\rm E}[\left\|  \B{x}^2 \right\|] \leq P_{\rm Tr}$. The vector $\B{\eta}$ represents the additive noise, whose entries are i.i.d. and distributed as $\mathcal{CN}(0,\sigma_\eta^2 )$.  The quantized channel output $\B{r}\in\mathbb{C}^{N}$ is thus represented as
\begin{equation}
 \B{r} = Q(\B{y}) = Q(\B{H}\B{x}+\B{\eta}).
\end{equation}
 In a one-bit system, the real parts $y_{i,R}$ and the imaginary parts $y_{i,I}$ of the unquantized receive signals $y_i$, $1\leq i\leq N$, are each quantized by a symmetric one-bit quantizer. Thus, the resulting quantized signals read as
 \begin{equation}
r_{i,c}=Q(y_{i,c})=\textrm{sign}(y_{i,c})=\left\{
\begin{array}{ll}
	+1 & {\rm if~}  y_{i,c}\geq 0 \\
	-1 & {\rm if~}  y_{i,c}< 0
\end{array}\right.
,\textrm{ for }c\in\{R,I\},\textrm{ }1 \leq i \leq  N,
\end{equation}
The operator $Q(\B{y})$ will also be denoted as ${\rm sign}(\B{y}) $ and represents the one-bit symmetric scalar quantization process in each real dimension. The restriction to one-bit symmetric quantization is motivated by its simple implementation. Since all of the real and imaginary components of the receiver noise $\B{\eta}$ are statistically independent with variance $\sigma_\eta^2$, we can express each of the conditional probabilities as the product of the conditional probabilities on each receiver dimension
\begin{equation}
\begin{aligned}
P(\B{r}={\rm sign}(\B{y})|\B{x},\B{H})&=\prod_{c\in\{R,I\}}\prod_{i=1}^{N}P(r_{i,c}|\B{x})\\
&=\prod_{c\in\{R,I\}}\prod_{i=1}^{N} \Phi\left( \frac{r_{i,c}[\B{H}\B{x}]_{i,c}}{\sqrt{\sigma_\eta^2/2}}  \right),
\label{cond_pro}
\end{aligned}
\end{equation}
where $\Phi(x) = \frac{1}{\sqrt{2\pi}}\int_{-\infty}^{x}e^{-\frac{t^2}{2}}dt$ is the cumulative normal distribution function.
We first state the main theorem used throughout the paper, and then provide the asymptotics of the mutual information for several channel models up to second order in the SNR.

\vspace{-0.2cm}

\section{Main theorem for the asymptotic entropy of one-bit quantized vector signals}
\label{entropy_theorem}
We provide a theorem that can be used for deriving the second order approximation of the mutual information. It considers the 1-bit signal $\B{r}={\rm sign}(\varepsilon \B{x} + \B{\eta} )$, where  $\B{x}$ is a random vector with a certain distribution and $\B{\eta}$ is random with i.i.d. Gaussian entries and unit variance, while $\varepsilon$ is a signal scaling parameter.
\newline
\newtheorem {theorem1}{Theorem}
\begin{theorem1}
\label{low_snr_mi}
Assuming $\B{x}\in \mathbb{C}^N$ is a proper complex random vector (${\rm E}[\B{x}\B{x}^{\rm T}]= {\rm E}[\B{x}]{\rm E}[\B{x}^{\rm T}]$) satisfying  $\textrm{E}_{\B{x}} [\left\|\B{x}\right\|_4^{4+\varepsilon}]<\delta$ for some finite constants $\varepsilon,\delta>0$ and $\B{\eta}$ is i.i.d. Gaussian with unit variance, then the following entropy approximation holds up to the second order in $\varepsilon^2$  
\begin{equation}
\begin{aligned}
H({\rm sign}(\varepsilon \B{x}+ \B{\eta}))  = & 2N\ln 2 -\frac{2}{\pi}\varepsilon^2 \left\|{\rm E}[\B{x}]\right\|_2^2-\varepsilon^4\left(  \frac{2}{\pi^2} {\rm tr}\left(\left({\rm nondiag}(\B{C}_x) \right)^2\right) \right.  \\ 
& \left. -\frac{4}{3\pi}\left\|{\rm E}[\B{x}]\circ {\rm E}[\B{x}\circ\B{x}\circ \B{x}] \right\|_1^1+\frac{4}{3\pi^2}\left\|{\rm E}[\B{x}]\right\|_4^4\right) + o(\varepsilon^4),
\end{aligned}
\end{equation}	
where $\left\|\B{a}\right\|_p^p=\sum_{i,c} |a_{i,c}|^p$ and $[\B{a}\circ\B{b}]_i=a_{i,R}b_{i,R}+{\rm j}a_{i,I}b_{i,I}$, while the expectation is taken with respect to $\B{x}$ and $\B{C}_x={\rm E}[\B{x}\B{x}^{\rm H}] - {\rm E}[\B{x}]{\rm E}[\B{x}^{\rm H}]$ is the covariance matrix of $\B{x}$.
\end{theorem1}

\begin{proof}
See appendix~\ref{proof_theorem1}.
\end{proof}
From this theorem, we can deduce some useful lemmas.

\newtheorem {lemma1}{Lemma}
\begin {lemma1}
  \label{lemma1_1}
For any possibly non-deterministic function $\B{g}(\B{x})$ satisfying  $\textrm{E}_{\B{x}} [\left\|\B{g}(\B{x})\right\|_4^{4+\varepsilon}]<\delta$ and ${\rm E}[\B{g}(\B{x})\B{g}(\B{x})^{\rm T}]= {\rm E}[\B{g}(\B{x})]{\rm E}[\B{g}(\B{x}^{\rm T})]$ for some finite constants $\varepsilon,\delta>0$, we have to the second order in $\varepsilon^2$
\begin{equation}
 \begin{aligned}
H({\rm sign}(\varepsilon \B{g}(\B{x})+ \B{\eta}))\approx & 2N\ln 2 -\frac{2}{\pi}\varepsilon^2 \left\|{\rm E}[\B{g}(\B{x})]\right\|_2^2
-\varepsilon^4\left(  \frac{2}{\pi^2} {\rm tr}\left(\left({\rm nondiag}(\B{C}_{\B{g}(\B{x})})) \right)^2\right) \right.  \nonumber \\
 &\left.-\frac{4}{3\pi}\left\|{\rm E}[\B{g}(\B{x})]\circ {\rm E}[\B{g}(\B{x})\circ\B{g}(\B{x})\circ \B{g}(\B{x})] \right\|_1^1+\frac{4}{3\pi^2}\left\|{\rm E}[\B{g}(\B{x})]\right\|_4^4\right).
 \end{aligned}
\end{equation}
\end{lemma1}

\newtheorem{lemma22}[lemma1]{Lemma}
\begin{lemma22}
 \label{lemma22_2}
For any function $\B{g}(\B{x})$ satisfying  $\textrm{E}_{\B{x}} [\left\|\B{g}(\B{x})\right\|_4^{4+\varepsilon}]<\delta$ and ${\rm E}[\B{g}(\B{x})\B{g}(\B{x})^{\rm T}|\B{x}]= {\rm E}[\B{x}|\B{x}]{\rm E}[\B{g}(\B{x})(\B{x})^{\rm T}|\B{x}]$, we have the following second order approximation of the conditional entropy

\begin{equation}
  \begin{aligned}
 H({\rm sign}(\varepsilon \B{g}(\B{x})+ \B{\eta})|\B{x}) \approx & 2N\ln 2 -\frac{2}{\pi}\varepsilon^2 {\rm E}_{\B{x}} [\left\|{\rm E}[\B{g}(\B{x})|\B{x}]\right\|_2^2]
-\varepsilon^4{\rm E}_{\B{x}}  \!\! \left[  \frac{2}{\pi^2} {\rm tr}\left(\left({\rm nondiag}(\B{C}_{\B{g}(\B{x})|\B{x}}) \right)^2\right) \right.  \nonumber \\
 &\left.-\frac{4}{3\pi}\left\|{\rm E}[\B{g}(\B{x})|\B{x}]\circ {\rm E}[\B{g}(\B{x}) \circ \B{g}(\B{x}) \circ \B{g}(\B{x})|\B{x}] \right\|_1^1+\frac{4}{3\pi^2}\left\|{\rm E}[\B{g}(\B{x})|\B{x}]\right\|_4^4\right].
  \end{aligned}
 \end{equation}

\end{lemma22}
\begin{proof}
 Lemma~\ref{lemma1_1}  is a direct result of Theorem~\ref{low_snr_mi}, where we just replace the random vector $\B{x}$ by $\B{g}(\B{x})$ if the stated conditions are fulfilled for $\B{g}(\B{x})$.  For Lemma~\ref{lemma22_2} we just perform the expectation in Theorem~\ref{low_snr_mi} first conditioned on $\B{x}$, to get the entropy for a given $\B{x}$ and then we take the average with respect to  $\B{x}$, again if the stated assumptions regarding the distribution of $\B{g}(\B{x})$ are fulfilled. These results will be used to derive a second order approximation of the mutual information of quantized MIMO systems for the case of perfect as well as statistical channel state information.    
\end{proof}

\section{Mutual Information and Capacity with full CSI}
\label{section:mutual}
When the channel $\B{H}$ is perfectly known at the receiver, the mutual information (in nats/s/Hz) between the channel input and the quantized output in Fig.~\ref{downlink_figure} reads as \cite{cover}
\begin{equation}
\begin{aligned}	
I(\B{x},\B{r})=\textrm{E}_{\B{x}}\left[\sum_{\B{r}}P(\B{r}|\B{x},\B{H})\textrm{ln}\frac{P(\B{r}|\B{x},\B{H})}{P(\B{r}|\B{H})}\right],
\label{transinfo}
\end{aligned}
\end{equation}
with $P(\B{r}|\B{H})=\textrm{E}_{\B{x}}[P(\B{r}|\B{x},\B{H})]$ and $\textrm{E}_{\B{x}}[\cdot]$ is the expectation taken with respect to $\B{x}$. For large $N$, the computation of the mutual information has intractable complexity due to the summation over all possible $\B{r}$, except for low dimensional outputs (see \cite{hea14} for the single output case), which is not relevant for the massive MIMO case. Therefore, we resort to a low SNR approximation to perform the analysis on the achievable rates.
\subsection{Second-order Expansion of the Mutual Information with 1-bit Receivers}
\label{section:duality}
In this section, we will elaborate on the second-order expansion of the
input-output mutual information (\ref{transinfo}) of the considered system in Fig.~\ref{downlink_figure} as
the signal-to-noise ratio goes to zero.\\
\newtheorem {theorem11}[theorem1]{Theorem}
\begin {theorem11}
\label{dualitytheorem}
Consider the one-bit quantized MIMO system in Fig.~\ref{downlink_figure} under a zero-mean input distribution $p(\B{x})$  with  covariance matrix $\B{C}_x$, satisfying $p(\B{x})=p(\textrm{j}\B{x}), \forall \B{x} \in \mathbb{C}^{M}$  (zero-mean proper complex distribution)\footnote{This restriction is simply justified by symmetry considerations.} and $\textrm{E}_{\B{x}} [\left\| \B{H}\B{x}\right\|_4^{4+\alpha}]<\gamma$ for some finite constants $\alpha,\gamma>0$. Then, to the second order, the mutual information (in nats) between the inputs and the quantized outputs with perfect CSI is given by:
\begin{equation}
\begin{aligned}	 
I(\B{x},\B{r})=&\frac{2}{\pi}\textrm{tr}(\B{H}\B{C}_x\B{H}^\textrm{H}) \frac{1}{\sigma_\eta^2}\!-\!\!\left[\frac{2}{\pi^2}\textrm{tr}((\textrm{nondiag}(\B{H}\B{C}_x\B{H}^\textrm{H}))^2)\right.\\
&\left.	+\frac{4}{3\pi}(1-\frac{1}{\pi}) \textrm{E}_{\B{x}} [\left\| \B{H}\B{x} \right\|_4^4]
\right]\frac{1}{\sigma_\eta^4}+\underbrace{\Delta I(\B{x},\B{r})}_{o(\frac{1}{\sigma_\eta^4})},
\label{transinfo_2order}
\end{aligned}
\end{equation}
where $\left\|\B{a} \right\|_4^4$ is the 4-norm of $\B{a}$ taken to the power 4: $\sum_{i,c}a_{i,c}^4$.\\
\end {theorem11}
\begin{proof}
We start with the definition of the mutual information \cite{cover}

\begin{equation}
\begin{aligned}
I(\B{x},\B{r}) &= H(\B{r}) - H(\B{r}|\B{x})  \\
               &= H({\rm sign}(\B{H}\B{x}+\B{\eta})|\B{H})  - H({\rm sign}(\B{H}\B{x}+\B{\eta})|\B{x},\B{H}),
\end{aligned}
\end{equation}
then we use Lemma~\ref{lemma1_1} and Lemma~\ref{lemma22_2} with $\varepsilon=\frac{1}{\sigma_\eta}$ and $\B{g}(\B{x})=\B{H}\B{x}$ to get the following asymptotic expression: 
\begin{equation}
\begin{aligned}
I(\B{x},\B{r})  &=    2N\ln 2 -\frac{2}{\pi}\frac{1}{\sigma_\eta^2} \left\|{\rm E}[\B{H}\B{x}]\right\|_2^2
-\frac{1}{\sigma_\eta^4}\left(  \frac{2}{\pi^2} {\rm tr}\left(\left({\rm nondiag}(\B{C}_{\B{H}\B{x}}) \right)^2\right) \right.  \nonumber \\
 &~~ \left.-\frac{4}{3\pi}\left\|{\rm E}[\B{H}\B{x}]\circ {\rm E}[\B{H}\B{x}\circ\B{H}\B{x}\circ \B{H}\B{x}] \right\|_1^1+\frac{4}{3\pi^2}\left\|{\rm E}[\B{H}\B{x}]\right\|_4^4\right)  \\
 &~~ - 2N\ln 2 +\frac{2}{\pi} \frac{1}{\sigma_\eta^2} {\rm E}_{\B{x}} [\left\|{\rm E}[\B{H}\B{x}|\B{x}]\right\|_2^2]
+ \frac{1}{\sigma_\eta^4} {\rm E}_{\B{x}}  \!\! \left[  \frac{2}{\pi^2} {\rm tr}\left(\left({\rm nondiag}(\B{C}_{\B{H}\B{x}|\B{x}}) \right)^2\right) \right.  \nonumber \\
 &~~\left.-\frac{4}{3\pi}\left\|{\rm E}[\B{H}\B{x}|\B{x}]\circ {\rm E}[\B{H}\B{x} \circ \B{H}\B{x} \circ \B{H}\B{x}|\B{x}] \right\|_1^1+ \frac{4}{3\pi^2}\left\|{\rm E}[\B{H}\B{x}|\B{x}]\right\|_4^4\right] +o(\frac{1}{\sigma_\eta^4})  \\
 &=   \frac{2}{\pi}\frac{1}{\sigma_\eta^2}  {\rm E} \left\|[\B{H}(\B{x}-{\rm E}[\B{x}])]\right\|_2^2
-\frac{1}{\sigma_\eta^4}\left(  \frac{2}{\pi^2} {\rm tr}\left(\left({\rm nondiag}(\B{H}\B{C}_x\B{H}^{\rm H}) \right)^2\right) \right.  \nonumber \\
 &~~\left.-\frac{4}{3\pi}\left\|{\rm E}[\B{H}\B{x}]\circ {\rm E}[\B{H}\B{x}\circ\B{H}\B{x}\circ \B{H}\B{x}] \right\|_1^1+\frac{4}{3\pi^2}{\rm E}[\left\|\B{H} {\rm E}[\B{x}] \right\|_4^4 ] \right.  \\
 &~~ \left.  + \frac{4}{3\pi}  {\rm E}[\left\|  \B{H}  \B{x} \right\|_4^4]  -\frac{4}{3\pi^2}{\rm E}[\left\|\B{H}\B{x}\right\|_4^4 ]   \right)  +o(\frac{1}{\sigma_\eta^4}).
\end{aligned}
\end{equation}
In the case that the distribution is zero-mean ${\rm E}[\B{x}]=\B{0}$, we end up exactly with the formula stated by the theorem.  
The condition $\textrm{E}_{\B{x}} [\left\|\B{H}\B{x}\right\|_4^{4+\alpha}]<\gamma$ for some finite constants $\alpha,\gamma>0$ is necessary, so that the remainder term of the expansion given by
\begin{equation}
\Delta I(\B{x},\B{r})=\textrm{E}_{\B{x}}[o(\left\|\B{H}\B{x}\right\|_4^4 \frac{1}{\sigma_\eta^4} )]
\end{equation}
satisfies
\begin{equation} 
 \lim_{ \frac{1}{\sigma^4}  \rightarrow 0}\frac{\Delta I(\B{x},\B{r})}{\frac{1}{\sigma_\eta^4}}=0,
 \end{equation} 
as already stated by Theorem~\ref{low_snr_mi}.
\end {proof}

For comparison, we use the results of Prelov and Verd\'{u} \cite{prelov} to express the mutual information  (in nats) between the  input $\B{x}$ and the unquantized output $\B{r}$ with the same  input distribution as in Theorem~\ref{dualitytheorem}:
\begin{equation}
\begin{aligned}	
I(\B{x},\B{y})=\textrm{tr}(\B{H}\B{C}_x\B{H}^\textrm{H}) \frac{1}{\sigma_\eta^2} -&\frac{\textrm{tr}((\B{H}\B{C}_x\B{H}^\textrm{H})^2)}{2} \frac{1}{\sigma_\eta^4} +o(\frac{1}{\sigma_\eta^4} ).
\label{transinfo_2order_unq}
\end{aligned}
\end{equation}
While  the mutual information for the unquantized channel in (\ref{transinfo_2order_unq}), up to the second order,  depends only on the input covariance matrix, in the quantized case (\ref{transinfo_2order}) it also depends on the fourth order statistics of $\B{x}$ (the fourth mixed moments of its components). \\

Now, using (\ref{transinfo_2order}) and (\ref{transinfo_2order_unq}), we deduce the mutual information penalty in the low SNR (or large dimension) regime incurred by quantization
\begin{equation}
\lim_{\frac{1}{\sigma_\eta^2} \rightarrow 0} \frac{I(\B{x},\B{r})}{I(\B{x},\B{y})}=\frac{2}{\pi},
\end{equation}
which is independent of the channel and the chosen distribution. Since the Gaussian distribution achieves the capacity of the unquantized channel but not necessarily for the  quantized case, we obtain for the supremum of the mutual information, i.e the capacity
\begin{equation}
\lim_{\frac{1}{\sigma_\eta^2} \rightarrow 0} \frac{C_\textrm{1-bit}}{C_\textrm{$\infty$-bit}} \geq \frac{2}{\pi}.
\end{equation}
These results can be also obtained based on the pseudo-quantization noise model \cite{mezghani2007,mezghani_2012_isit} and it generalizes the result known for the AWGN channel \cite{Viterbi}. \\

\begin{figure}[h]
\begin{center}
\psfrag{Mutual Information in bits}[c][c]{\small Mutual Information in bpcu}
\psfrag{SNR (linear)}[c][c]{\small SNR (linear)}
{\epsfig{file=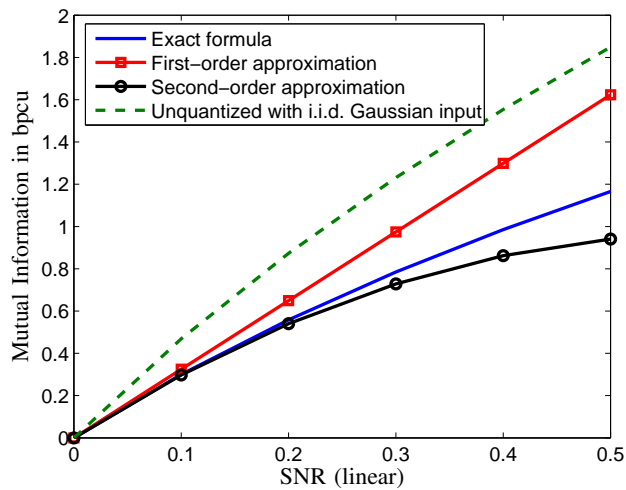, width = 9.2cm}}
\caption{Mutual information of a 1-bit quantized 4$\times$4 QPSK MIMO system and its first and second-order approximations. For comparison the mutual information without
quantization is also plotted.}
\label{transinfo_fig}
\end{center}
\end{figure}

Fig.~\ref{transinfo_fig} illustrates the mutual information for a randomly generated 4$\times$4  channel\footnote{The generated entries $h_{i,j}$ of $\B{H}$ are uncorrelated and $h_{i,j}\sim\mathcal{CN}(0,1).$} with QPSK signaling and total power ${\rm tr}(\B{C}_x)=P_{\rm Tr}={\rm SNR} \cdot \sigma_\eta^2$, computed exactly using (\ref{transinfo}), and also its first and second-order approximations from (\ref{transinfo_2order}). For comparison, the mutual information without quantization (using i.i.d. Gaussian input) is also plotted. Fig.~\ref{transinfo_fig} shows that the ratio $\frac{2}{\pi}$ holds for low to moderate $\textrm{SNR}=\frac{P_{\rm Tr}}{\sigma_\eta^2}$.    \\

For a larger number of antennas, the inner summation in (\ref{transinfo}) may be intractable. In this case, the second-order approximation  in (\ref{transinfo_2order}) is advantageous at low SNR to overcome the high complexity of the exact formula.
\subsection{Capacity with Independent-Component Inputs}
\label{section:receiver}
Lacking knowledge of the channel or its statistics, the transmitter assigns the power evenly over the  components $x_{i,c}$ of the input vector $\B{x}$, i.e., $\textrm{E}[x_{i,c}^2]=\frac{P_{\rm Tr}}{2M}$, in order to achieve good performance on average. Furthermore, let us assume these components to be independent of each other (e. g. multi-streaming scenario).\footnote{Clearly, this is not necessarily the capacity achieving strategy.} Thus, the probability density function  of the input vector $\B{x}$ is $p(\B{x})=\prod_{i,c}p_ {i,c}(x_{i,c})$.\footnote{Note that the $p_ {i,c}(x_{i,c})$ have to be even functions and $p_ {i,R}(x_{i,R})=p_ {i,I}(x_{i,I})~\forall i$, due to the symmetry (see Theorem~\ref{dualitytheorem}) and convexity of the mutual information.} Now, with
\begin{eqnarray}
[\B{H}\B{x}]_{i,R}&=&\left( \sum_j [h_{i,j,R}x_{j,R}-h_{i,j,I}x_{j,I} ] \right),\\
\mu_{j,c}&=&\frac{\textrm{E}[x_{j,c}^4]}{\textrm{E}[x_{j,c}^2]^2}=4M^2\textrm{E}[x_{j,c}^4],
\end{eqnarray}
and the \emph{kurtosis} of the random component $x_{j,c}$ defined as 
\begin{equation}
\kappa_{j,c}= \mu_{j,c}-3,
\end{equation}
we get
\begin{equation}
\begin{aligned}	
\textrm{E}_{\B{x}}[([\B{H}\B{x}]_{i,R})^4]\!&=\!\frac{1}{4M^2}\!\left(3\!\!\!\!\!\!\sum_{\stackrel{j,c,j',c'}{(j,c)\neq(j',c')}}\!\!\!\!\!\!\!\!\!h_{i,j,c}^2h_{i,j',c'}^2\!+\!\sum_{j,c}\mu_{j,c}h_{i,j,c}^4  \right) \\
&=\frac{1}{4M^2}\!\left(3 \left(\left[\B{H}\B{H}^\textrm{H}\right]_{i,i}\right)^2 +\sum_{j,c}\kappa_{j,c}h_{i,j,c}^4 \right).
\end{aligned}	
\end{equation}

Similar results hold for the other components of the vector $\B{H}\B{x}$. Plugging this result and $\B{C}_x= \frac{P_{\rm Tr}}{M}\textbf{I}$ into (\ref{transinfo_2order}), we obtain an expression for the mutual information with independent-component inputs and $\B{C}_x= \frac{P_{\rm Tr}}{M}\textbf{I}$ up to second order:
\begin{equation}
\begin{aligned}	
&I^{\textrm{ind}}(\B{x},\B{r})\approx\frac{2}{\pi}\textrm{tr}(\B{H}\B{H}^\textrm{H})\frac{P_{\rm Tr}}{M \sigma_\eta^2}\!-\!\!\Bigg[\frac{2}{\pi^2}\textrm{tr}((\textrm{nondiag}(\B{H}\B{H}^\textrm{H}))^2)\Big.\\
&\!\left.	+\frac{2}{3\pi}(1\!-\!\frac{1}{\pi})\!\! \left( 3~ \textrm{tr}((\textrm{diag}(\B{H}\B{H}^\textrm{H}))^2)+\!\sum_{i,j,c}\kappa_{j,c} h_{i,j,c}^4 \!\!\right)
\right]\!\!  \left(\frac{P_{\rm Tr}}{M \sigma_\eta^2} \right)^2.
\label{transinfo_ind_2order}
\end{aligned}
\end{equation}
Now,  we state a theorem on the structure of the near-optimal input distribution under these assumptions.\\
\newtheorem {theorem3}[theorem1]{Theorem}
\begin{theorem3}
\label{opt_dheorem}
To second order, QPSK signals are capacity-achieving among all signal distributions with independent components. The achieved capacity up to second order is
\begin{equation}
\begin{aligned}	
&C_\textrm{1-bit}\approx\frac{2}{\pi}\textrm{tr}(\B{H}\B{H}^\textrm{H})\frac{P_{\rm Tr}}{M \sigma_\eta^2}-\Bigg[\frac{2}{\pi^2}\textrm{tr}((\textrm{nondiag}(\B{H}\B{H}^\textrm{H}))^2)\Big.\\
&\!\left.	+\frac{2}{3\pi}\!(1\!-\!\frac{1}{\pi}) \left( 3~ \textrm{tr}((\textrm{diag}(\B{H}\B{H}^\textrm{H}))^2)-2\sum_{i,j,c} h_{i,j,c}^4 \!\!\right)
\right]  \left( \frac{P_{\rm Tr}}{M \sigma_\eta^2}  \right)^2.
\label{capacity_2order}
\end{aligned}
\end{equation}
\end{theorem3}
\vspace{0.3cm}
\begin{proof}
Since $\textrm{E}[x_{i,c}^4]\geq \textrm{E}[x_{i,c}^2]^2$, we have $\kappa_{i,c}=\frac{\textrm{E}[x_{i,c}^4]}{\textrm{E}[x_{i,c}^2]^2}-3  \geq -2,\forall i,c$.  Obviously, the QPSK distribution is the unique distribution with independent-component inputs that can achieve all these lower bounds simultaneously, i.e., $\kappa_{i,c}=\kappa_{\textrm{QPSK}}=-2~\forall i,c$, and thus maximize $I^{\textrm{ind}}(\B{x},\B{r})$ in (\ref{transinfo_ind_2order}) up to second order. 
\end{proof}
\subsection{Ergodic Capacity under i.i.d. Rayleigh Fading Conditions}
\label{section:capacity}
Here we assume the channel $\B{H}$ to be ergodic with i.i.d. Gaussian components $h_{i,j}\sim\mathcal{CN}(0,1)$. The ergodic capacity can be written as
  \begin{equation}
  C^{\textrm{erg}}_{\textrm{1-bit}}=\textrm{E}_{\B{H}}[C_{\textrm{1-bit}}].
    \end{equation}
    We apply the expectation over $\B{H}$ using the second order expansion of $C_{\textrm{1-bit}}$ in (\ref{capacity_2order}). By expanding the following expressions and taking the expectation over the i.i.d. channel coefficients, we have  
\begin{eqnarray}
	 	\textrm{E}_{\B{H}}\left[\textrm{tr}(\B{H}\B{C}_x\B{H}^{\textrm{H}})\right]&=& N \textrm{tr}(\B{C}_x)\\
	 	\textrm{E}_{\B{H}}\left[\textrm{tr}\left((\textrm{nondiag}(\B{H} \B{C}_x  \B{H}^{\textrm{H}}))^2\right)\right]&=& N(N-1)\textrm{tr}(\B{C}_x^2) \label{EH2}\\
	 	\textrm{E}_{\B{H}} \left[\left\| \B{H}\B{x} \right\|_4^4\right] &=& \frac{3}{2}N\textrm{E}_{\B{x}} \left[\left\| \B{x} \right\|_2^4 \right].
\end{eqnarray}
The ergodic mutual information over an i.i.d. channel can be obtained as
\begin{equation}
  \textrm{E}_{\B{H}}[I(\B{x},\B{r})] \approx N\textrm{tr}(\B{C}_x)\frac{2}{\pi}\frac{1}{\sigma_\eta^2}-\frac{N}{2}\left((N-1)\textrm{tr}(\B{C}_x^2)+(\pi-1)\textrm{E}_{\B{x}} \left[\left\| \B{x} \right\|_2^4\right]\right)\left(\frac{2}{\pi} \frac{1}{\sigma_\eta^2} \right)^2.
\end{equation}
Next, we characterize the capacity achieving distribution up to second order in the SNR.
\newtheorem {theorem_erg}[theorem1]{Theorem}
\begin{theorem_erg}
\label{theorem_erg_cap}
The ergodic capacity of the 1-bit quantized i.i.d. MIMO channel is achieved asymptotically at low SNR by QPSK signals and reads as 
\begin{equation}
  C^{\textrm{erg}}_{\textrm{1-bit}}\approx N\frac{2}{\pi} \frac{P_{\rm Tr}}{\sigma_\eta^2} - \frac{N(N+(\pi-1)M-1)}{2M}\left(\frac{2}{\pi} \frac{P_{\rm Tr}}{\sigma_\eta^2} \right)^2.
  \label{C_erg_Q}
\end{equation}
\end{theorem_erg}
\begin{proof}
Since $\textrm{tr}(\B{C}_x^2)/M \geq (\textrm{tr}(\B{C}_x)/M)^2={P_{\rm Tr}/M}^2$ with equality if $\B{C}_x=\frac{P_{\rm Tr}}{M}\textbf{I}$, and $\textrm{E}_{\B{x}} \left[\left\| \B{x} \right\|_2^4 \right] \geq \left(\textrm{E}_{\B{x}} \left[\left\| \B{x} \right\|_2^2\right]\right)^2=P_{\rm Tr} $ with equality if the input has a constant norm of 1, the ergodic capacity is achieved by a constant-norm uncorrelated input ($\B{C}_x=\frac{P_{\rm Tr}}{M} \textbf{I} $ and $\left\| \B{x} \right\|_2^2=P_{\rm Tr}$), which can be obtained for instance by QPSK signals
Finally, using (\ref{capacity_2order}) we obtain the second-order expression for the ergodic capacity of one-bit quantized Rayleigh fading channels for the  QPSK case.
\end{proof}
Compared to the ergodic capacity in the unquantized case achieved by i.i.d. Gaussian inputs (or even by QPSK up to the second order)  \cite{verdu}
 \begin{equation}
  C^{\textrm{erg}}\approx N\cdot \frac{P_{\rm Tr}}{\sigma_\eta^2} -\frac{N(N+M)}{2M}  \left(\frac{P_{\rm Tr}}{\sigma_\eta^2} \right)^2,
  \label{C_erg}
\end{equation}
the ergodic capacity of one-bit quantized MIMO under QPSK $C^{\textrm{erg}}_{\textrm{1-bit}}$ incorporates 
a power penalty of almost $\frac{\pi}{2}$ (1.96 dB), when considering only the linear term that characterizes the capacity in the limit of infinite bandwidth.\\

On the other hand, the second order term quantifies the convergence of the capacity function to the low SNR limit, i.e. the first order term,  by reducing the power or increasing the bandwidth \cite{verdu}. Therefore, it can be observed from
\begin{equation}
1<\frac{N+(\pi-1)M-1}{N+M}<\pi-1,
\end{equation}
 that the quantized channel converges to this limit slower than the unquantized channel. Nevertheless, for  $M=1$ or $M \ll N$ (massive MIMO uplink scenario), this difference in the convergence behavior vanishes almost completely, since both second-order expansions (\ref{C_erg_Q}) and (\ref{C_erg}) become nearly the same up to the factor $2/\pi$ in \textrm{SNR}. \\

\begin{figure}[h]
\begin{center}
\psfrag{M}[c][c]{$M$}
\psfrag{N=20}[c][c]{$N\!=\!20$}
\psfrag{N=30}[c][c]{$N\!=\!30$}
\psfrag{N=40}[c][c]{$N\!=\!40$}
\psfrag{M1/M}[c][c]{$M_{\rm 1-bit}/M$}
{\epsfig{file=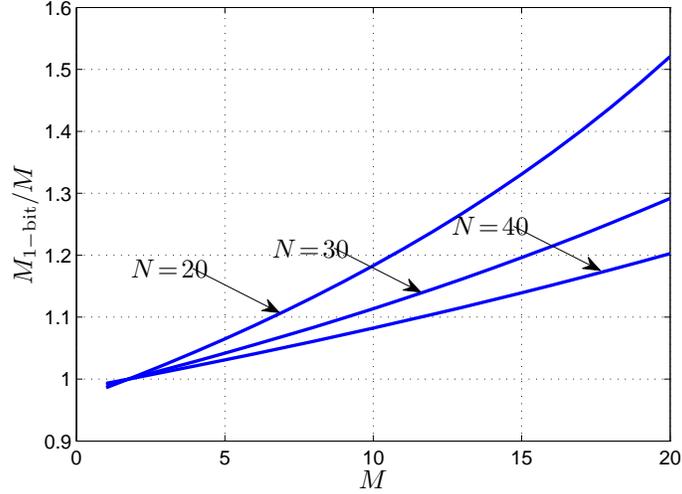, width = 9.2cm}}
\caption{Required $M_{\rm 1-bit}/M$ to achieve ideal ergodic capacity up to the second order.}
\label{M1vsM}
\end{center}
\end{figure}

In addition, the ergodic capacity of the quantized channel $C^{\textrm{erg}}_{\textrm{1-bit}}$ in (\ref{C_erg_Q}) increases linearly with the number of receive antennas $N$ and only  sublinearly with the number of transmit antennas $M$, which holds also for  $C^{\textrm{erg}}$. For the special case of one receive antenna, $N=1$, $C^{\textrm{erg}}_{\textrm{1-bit}}$ does not depend on the number of transmit antennas $M$ up to the second order, contrary to  $C^{\textrm{erg}}$. On the other hand, if one would achieve, up to the second order, the same ergodic capacity at the same power with one-bit receivers as in the ideal case by adjusting the number of receive and transmit antennas, i.e.,
 \begin{equation}
  N\cdot \frac{P_{\rm Tr}}{\sigma_\eta^2} -\frac{N(N+M)}{2M}  \left(\frac{P_{\rm Tr}}{\sigma_\eta^2} \right)^2=N_{\rm 1-bit}\frac{2}{\pi} \frac{P_{\rm Tr}}{\sigma_\eta^2} - \frac{N_{\rm 1-bit}(N_{\rm 1-bit}+(\pi-1)M_{\rm 1-bit}-1)}{2M_{\rm 1-bit}}\left(\frac{2}{\pi} \frac{P_{\rm Tr}}{\sigma_\eta^2} \right)^2,
 \end{equation}
 then we can deduce by equating coefficients that
\begin{equation}
\begin{aligned}	
N_{\rm 1-bit}&= \frac{\pi}{2} N, \\
M_{\rm 1-bit}&= M \frac{\pi N-2}{\pi N -(\pi-2)M}.
\end{aligned}	
\end{equation}
The one-bit receive dimension has to be increased by $\pi/2$, while the behavior for the number of transmit antennas is shown in Fig.~\ref{M1vsM}. Clearly, when $N \gg M$, which corresponds to a typical massive MIMO uplink scenario, we have $M_{\rm 1-bit} \approx M$. This means that, at the transmitter (or user) side, there is no need to increase the number of antennas up to the second order in SNR, showing that the total increase of dimensions is moderate.

\section{Mutual Information and Capacity with Statistical CSI and 1-bit Receivers}

  We reconsider now the extreme case of 1-bit quantized communications over MIMO Rayleigh-fading channels but assuming that  only the statistics of the channel are known at the receiver. Later, we will also treat the achievable rate for the SISO channel case for the whole SNR range.

\label{section:mutual_n}
Generally, the mutual information (in nats/s/Hz) between the channel input and the quantized output in Fig.~\ref{downlink_figure} with statistical CSI reads as \cite{cover}
\begin{equation}
\begin{aligned}	
I(\B{x},\B{r})&=H(\B{r})-H(\B{r}|\B{x})\\
&=\textrm{E}_{\B{x}}\left[\sum_{\B{r}}P(\B{r}|\B{x})\textrm{ln}\frac{P(\B{r}|\B{x})}{P(\B{r})}\right],
\label{transinfo_n}
\end{aligned}
\end{equation}
where $P(\B{r})=\textrm{E}_{\B{x}}[P(\B{r}|\B{x})]$ and $H(\cdot)$ and $H(\cdot|\cdot)$ represent the entropy and the conditional entropy, respectively. If the channel is Gaussian distributed with zero mean, then  
given the input $\B{x}$, the unquantized output $\B{y}$ is zero-mean complex Gaussian with covariance  $\textrm{E}[\B{y}\B{y}^{\rm H}|\B{x}]=\sigma_\eta^2 \cdot \textbf{I}_N+{\rm E}_{\B{H}}[\B{H}\B{x}\B{x}^{\rm H} \B{H}^{\rm H}] $, and thus we have   \\
\begin{equation}
   p(\B{y}|\B{x})=\frac{{\rm exp}(-\B{y}^{\rm H}(\sigma_\eta^2 \textbf{I}_N + \rm{E}_{\B{H}}[\B{H}\B{x}\B{x}^{\rm H} \B{H}^{\rm H}])^{-1}\B{y})}{\pi^{N}\left| \sigma_\eta^2 \textbf{I}_N + \rm{E}_{\B{H}}[\B{H}\B{x}\B{x}^{\rm H} \B{H}^{\rm H}] \right|}.
   \label{unq_cond_pro_n}
\end{equation}
Thus, we can express the conditional probability of the quantized output as
\begin{equation}
\begin{aligned}
P(\B{r}|\B{x}) & =\int_0^{\infty} \cdots \int_0^{\infty} p(\B{y} \circ \B{r}|\B{x}) {\rm d}\B{r} \\
&= \int_0^{\infty} \cdots   \int_0^{\infty}   \frac{{\rm exp}(-(\B{y} \circ \B{r} )^{\rm H}(\sigma_\eta^2 \cdot \textbf{I}_N + \rm{E}[\B{H}\B{x}\B{x}^{\rm H} \B{H}^{\rm H}])^{-1}(\B{y} \circ \B{r}))}{\pi^{N}\left|\sigma_\eta^2 \cdot \textbf{I}_N + \rm{E}[\B{H}\B{x}\B{x}^{\rm H} \B{H}^{\rm H}] \right|}{\rm d}\B{y},
\label{cond_pro_n}
\end{aligned}
\end{equation}
where the integration is performed over the positive orthant of the complex hyperplane. \\
The evaluation of this multiple integral is in general intractable. Thus, we consider first a  simple lower bound involving the mutual information under perfect channel state information at the receiver, which turns out to be tight in some cases as shown later. The lower bound is obtained by the chain rule and the non-negativity of the mutual information:
\begin{equation}
\begin{aligned}	
I(\B{x},\B{r})&=\textrm{E}_{\B{x},\B{H}}\left[\sum_{\B{r}}P(\B{r}|\B{x},\B{H})\textrm{ln}\frac{\textrm{E}_{\B{H}}[P(\B{r}|\B{x},\B{H})]}{\textrm{E}_{\B{x},\B{H}}[P(\B{r}|\B{x},\B{H})]}\right] \\
&=I(\B{H},\B{r})+I(\B{x},\B{r}|\B{H})-I(\B{H},\B{r}|\B{x})\\
&\geq I(\B{x},\B{r}|\B{H})-I(\B{H},\B{r}|\B{x})\\
&=\textrm{E}_{\B{x},\B{H}}\left[\sum_{\B{r}}P(\B{r}|\B{x},\B{H})\textrm{ln}\frac{\textrm{E}_{\B{H}}[P(\B{r}|\B{x},\B{H})]}{\textrm{E}_{\B{x}}[P(\B{r}|\B{x},\B{H})]}\right] \\
&=\textrm{E}_{\B{x},\B{H}}\left[\sum_{\B{r}}P(\B{r}|\B{x},\B{H})\textrm{ln}\frac{P(\B{r}|\B{x})}{P(\B{r}|\B{H})}\right].
\label{transinfo_lower}
\end{aligned}
\end{equation}
On the other hand, an upper bound is given by the coherent assumption (channel perfectly known at the receiver) 
\begin{equation}
\begin{aligned}	
I(\B{x},\B{r}) \leq  \textrm{E}_{\B{x},\B{H}}\left[\sum_{\B{r}}P(\B{r}|\B{x},\B{H}) \cdot \textrm{ln} \frac{P(\B{r}|\B{x},\B{H})}{P(\B{r}|\B{H})}\right],
\end{aligned}
\label{coh_siso}
\end{equation}
where we can express each of the conditional probabilities $P(\B{r}|\B{x},\B{H})$ as the product of the conditional probabilities on each receiver dimension, since the real and imaginary components of the receiver noise $\B{\eta}$ are statistically independent with power $\frac{1}{2}$ in each real dimension: 
\begin{equation}
\begin{aligned}
P(\B{r}|\B{x},\B{H})&=\prod_{c\in\{R,I\}}\prod_{i=1}^{N}P(r_{i,c}|\B{x})\\
&=\prod_{c\in\{R,I\}}\prod_{i=1}^{N}\Phi\left(r_{i,c}[\B{H}\B{x}]_{i,c}\sqrt{2\textrm{SNR}}\right),
\end{aligned}
\end{equation}
where $\Phi(x)=\frac{1}{\sqrt{2\pi}}\int_{-\infty}^{x}e^{-\frac{t^2}{2}}dt$ is the cumulative normal distribution function. Evaluating the lower bound in (\ref{transinfo_lower}), even numerically, is very difficult, except for some simple cases such as SISO block fading channels, as considered next. 
\subsection{The non-coherent block-Rayleigh fading SISO Case}
\label{section:siso_n}

Here we treat the block-Rayleigh fading SISO case in more detail, where $\B{H}=h\cdot\mathbf{I}_T$, $h \sim \mathcal{CN}(0,1)$, and $M=N=T$ is the coherence time. For simplicity and ease of notation we assume that $\sigma_\eta^2 = 1$, therefore we have without loss of generality $\textrm{SNR}=\frac{P_{\rm Tr}}{\sigma_\eta^2}= P_{\rm Tr} $.
The covariance matrix $\textrm{E}[\B{y}\B{y}^{\rm H}|\B{x}]=\textbf{I}_T+ \B{x}\B{x}^{\rm H}$ is the sum of an identity matrix and a rank one matrix. Then, we obtain the conditional probability of the 1-bit output as 
\begin{equation}
\begin{aligned}
\!\!\!P(\B{r}|\B{x})&={\rm E}_{h} [P( {\B{r}}| \B{x},h)]  \\
&= \frac{1}{\pi}  \int_{\mathbb{C}}   e^{-|h|^2} \prod_{t=1}^{T} \prod_{c \in \{R,I\}} \Phi(\sqrt{2}([h x_t]_{c})r_{t,c}) \cdot {\rm d}h \\
&=\frac{1}{\pi}\int_{\mathbb{C}} e^{-|h|^2}  \prod_{t=1}^{T} \Phi(\sqrt{2}{\rm Re}(hx_t)r_{t,R})\cdot \Phi(\sqrt{2}{\rm Im}(h x_t)r_{t,I}){\rm d}h \\
&=\frac{1}{2\pi}\int_{-\infty}^{+\infty}\!\!\!\int_{-\infty}^{+\infty}\!\!\!\!\!  e^{-\frac{u^2+v^2}{2}}  \prod_{t=1}^{T} \Phi( (x_{t,R}u-x_{t,I}v)r_{t,R})\cdot \Phi((x_{t,R}v+x_{t,I}u)r_{t,I}){\rm d}u{\rm d}v,
\label{cond_pro_siso}
\end{aligned}
\end{equation}
where $\Phi(x)=\frac{1}{\sqrt{2\pi}}\int_{-\infty}^{x}e^{-\frac{t^2}{2}}dt$ is the cumulative normal distribution function.

\subsubsection{Achievable Rate with i.i.d. QPSK for the 1-bit Block Fading SISO Model}
With the above formula, the achievable rate of the one-bit quantized SISO channel with QPSK input reads as
\begin{equation}
\begin{aligned}
 C^{\rm QPSK}_{\rm 1-bit}(\textrm{SNR})  &=\frac{1}{4^T} \sum_{\B{x}} \sum_{\B{r}}P(\B{r}|\B{x}) \log_2 ( \frac{P(\B{r}|\B{x})}{P(\B{r})} )\\
&= \sum_{\B{r}}P(\B{r}|\B{x}^{0}) \log_2 (4^T P(\B{r}| \B{x}^{0} ) ),
\end{aligned}
\label{R_QPSK}
\end{equation}
 Here, $\B{x}$ is drawn from all possible  sequences of $T$ equally likely QPSK data symbols, i.e., $x \in \{\sqrt{\textrm{SNR}},-\sqrt{\textrm{SNR}},{\rm j}\sqrt{\textrm{SNR}},-{\rm j}\sqrt{\textrm{SNR}} \}^T$ and $\B{x}^{0}$ denotes the constant sequence $x^{0}_t=\sqrt{\textrm{SNR}},\forall t$. The second equality in (\ref{R_QPSK}) follows due to the symmetry of the QPSK constellation and since
\begin{equation}
\begin{aligned}
P(\B{r})=\frac{1}{4^T} \sum\limits_{\B{x}} P(\B{r}|\B{x})=\frac{1}{4^T} \sum\limits_{\B{x}' \in\{\pm 1, \pm {\rm j}\}} P( \B{r}|\B{x}'\circ \B{x}^{0})\stackrel{\textrm{due to (\ref{cond_pro_siso})}}{=}\frac{1}{4^T} \sum\limits_{\B{x}' \in\{\pm 1, \pm {\rm j}\}}  P(\B{x}'\circ \B{r}|\B{x}^{0}) =\frac{1}{4^T} \sum\limits_{\B{r}'} P( \B{r}'|\B{x}^{0})=\frac{1}{4^T}.
\end{aligned}
\end{equation}
We note that the rate expression (\ref{R_QPSK}) corresponds exactly to its lower bound in (\ref{transinfo_lower}) due to the fact that in the i.i.d. QPSK case $P(\B{r}|h)=P(\B{r})=4^{-T}$ and thus $I(h,\B{r})=0$. Furthermore, we use (\ref{cond_pro_siso}) to get  a simpler expression for $P(\B{r}|\B{x}^{0})$ as
\begin{equation}
\begin{aligned}
P(\B{r}|\B{x}^{0}) & = \frac{1}{2\pi}\int_{-\infty}^{+\infty}\!\!\!\!\!e^{-\frac{u^2}{2}} \!\prod_{t=1}^{T}\! \Phi(\sqrt{\textrm{SNR}}r_{t,R}u){\rm d}u
\!\!\!\int_{-\infty}^{+\infty}\!\!\!e^{-\frac{v^2}{2}}\!\prod_{t=1}^{T}\!\Phi(\sqrt{\textrm{SNR}}r_{t,I}v) {\rm d}v\\
&=P({\rm Re}(\B{r})|\B{x}^{0}) \cdot P({\rm Im}(\B{r})|\B{x}^{0}).
\label{cond_pro_A}
\end{aligned}
\end{equation}
Then (\ref{R_QPSK}) simplifies to 
\begin{equation}
\begin{aligned}
 \!\!\! & C^{\rm QPSK}_{\rm 1-bit}(\textrm{SNR})= 2
\sum_{t,r_t=\pm 1}\left(\int_{-\infty}^{+\infty}\!\!\!\frac{e^{-\frac{u^2}{2}}}{\sqrt{2\pi}} \!\prod_{t=1}^{T}\! \Phi(\sqrt{\textrm{SNR}}r_{t}u){\rm d}u \right) \!\! \cdot  \log_2 \left(2^T  \int_{-\infty}^{+\infty}\!\!\!\frac{e^{-\frac{u^2}{2}}}{\sqrt{2\pi}} \!\prod_{t=1}^{T}\! \Phi(\sqrt{\textrm{SNR}}r_{t}u){\rm d}u   \right)\\
&= 2 \sum_{k=0}^{T} \left(\!\!
\begin{array}{c}
	T\\
	k
\end{array}
\!\! \right)  \left(\int_{-\infty}^{+\infty}\!\!\!\frac{e^{-\frac{u^2}{2}}}{\sqrt{2\pi}}  \Phi(-\sqrt{\textrm{SNR}}u)^k \Phi(\sqrt{\textrm{SNR}}u)^{T-k}  {\rm d}u \right) \!\! \cdot \log_2 \left(2^T \int_{-\infty}^{+\infty}\!\!\!\frac{e^{-\frac{u^2}{2}}}{\sqrt{2\pi}}  \Phi(-\sqrt{\textrm{SNR}}u)^k \Phi(\sqrt{\textrm{SNR}}u)^{T-k}   {\rm d}u  \right).
\end{aligned}
\label{R_QPSK2}
\end{equation}
For the special case $T=2$, 
we use the following closed form solution from \cite{bacon}  for the integral in (\ref{R_QPSK2}) 
\begin{equation}
\begin{aligned}
\frac{1}{\sqrt{2\pi}}\int_{-\infty}^{+\infty}\!\!\!e^{-\frac{u^2}{2}}\!&\prod_{t=1}^{2}\!\Phi(\sqrt{\textrm{SNR}}r_{t,c}u){\rm d}u= \frac{1}{4}\left[1+\frac{2}{\pi}\arcsin(\frac{\textrm{SNR}}{1+\textrm{SNR}})r_{1,c}r_{2,c} \right].
\end{aligned}
\end{equation}
Thus (\ref{R_QPSK2}) becomes
\begin{equation}
\begin{aligned}
 C^{\rm QPSK}_{{\rm 1-bit},T=2}(\textrm{SNR}) = &
  \left( 1+ \frac{2}{\pi} \arcsin(\frac{\textrm{SNR}}{1+\textrm{SNR}}) \right) \log_2
\left(1+ \frac{2}{\pi} \arcsin(\frac{\textrm{SNR}}{1+\textrm{SNR}}) \right)  +   \\
 & \left(1- \frac{2}{\pi} \arcsin(\frac{\textrm{SNR}}{1+\textrm{SNR}}) \right) \log_2 \left( 1- \frac{2}{\pi} \arcsin(\frac{\textrm{SNR}}{1+\textrm{SNR}}) \right). \\
\end{aligned}
\end{equation}
And for the case $T=3$, we substitute using again the corresponding formula from \cite{bacon})
\begin{equation}
\begin{aligned}
&\frac{1}{\sqrt{2\pi}}\int_{-\infty}^{+\infty}\!\!\!e^{-\frac{u^2}{2}}\!\prod_{t=1}^{3}\!\Phi(\sqrt{\textrm{SNR}}r_{t,c}u){\rm d}u=\\
&\frac{1}{8}\left[ 1 + \frac{2}{\pi}\left(\arcsin(\frac{\textrm{SNR}}{1+\textrm{SNR}})r_{1,c}r_{2,c}+\arcsin(\frac{\textrm{SNR}}{1+\textrm{SNR}})r_{1,c}r_{3,c} +\arcsin(\frac{\textrm{SNR}}{1+\textrm{SNR}})r_{2,c}r_{3,c}   \right) \right]
\end{aligned}
\end{equation}
 in (\ref{R_QPSK2}) and we obtain 
\begin{equation}
\begin{aligned}
C_{{\rm 1-bit},T=3}(\textrm{SNR})= & \left( \frac{1}{2}+ \frac{3}{\pi} \arcsin(\frac{\textrm{SNR}}{1+\textrm{SNR}}) \right) \log_2
\left(1+ \frac{6}{\pi} \arcsin(\frac{\textrm{SNR}}{1+\textrm{SNR}}) \right) +  \\
& \left(\frac{3}{2}- \frac{3}{\pi} \arcsin(\frac{\textrm{SNR}}{1+\textrm{SNR}}) \right) \log_2 \left(1- \frac{2}{\pi} \arcsin(\frac{\textrm{SNR}}{1+\textrm{SNR}}) \right). 
\label{rate_T3}
\end{aligned}
\end{equation}
 A closed form solution for the integral in (\ref{cond_pro_A}) for $T>3$ is unknown and only approximations are found in the literature \cite{bacon}.\\
  
\begin{figure}[h]
\begin{center}
\psfrag{C/T (bits/channel use)}[c][c]{$C/T$ (bits/channel use)}
\psfrag{SNR (linear)}[c][c]{$\textrm{SNR}$ (linear)}
{\epsfig{file=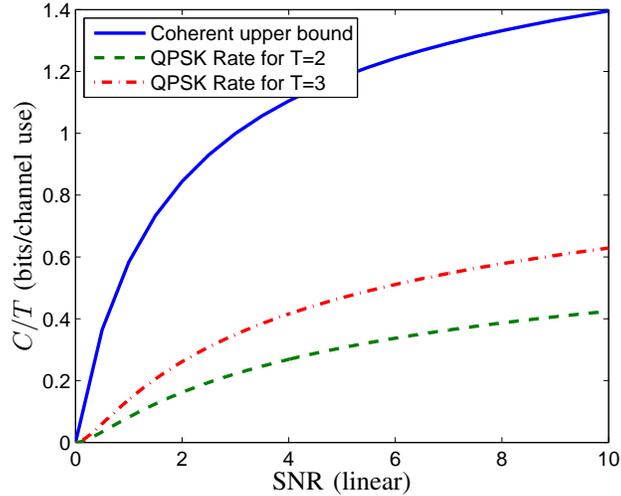, width = 8.7cm}}
\caption{Capacity of the one-bit Rayleigh-fading SISO channel for $T=2$, $T=3$, $T=\infty$.}
\label{transinfo_fig_n}
\end{center}
\end{figure}

In Fig.~\ref{transinfo_fig_n}, we plot the capacity of the one-bit SISO Rayleigh-fading channel per channel use  over the \textrm{SNR} for two cases $T=2$ and $T=3$. 
The coherent capacity achieved by quantized QPSK, which is an upper bound on our non-coherent capacity (see (\ref{coh_siso})), is also shown. It is obtained assuming that the receiver knows the channel coefficient $h$ (or $T$ goes to infinity). The average capacity achieved by QPSK over the quantized coherent Rayleigh-fading channel is obtained from (\ref{coh_siso}):
\begin{equation}
\begin{aligned}
  C^{\rm coh}_{\rm 1-bit}&=\textrm{E}_{h}[H(x,r|h)]\\
     &=\textrm{E}_{h}\left[2+\sum_x\sum_y P(r|x,h) \log_2 P(r|x,h)\right] \\
     &=2\left(1-\frac{1}{\sqrt{2\pi}} \int_{-\infty}^{\infty} {\rm e}^{-\frac{u^2}{2}}H_b(\Phi(\sqrt{ \textrm{SNR}} \cdot u )) {\rm d}u \right),
  \end{aligned}
\end{equation}
where we have used the binary entropy function $H_b(p)=-p \cdot \log_2 p - (1-p) \cdot \log_2 (1-p)$,  
and we substituted $H(r)=2$  since all four possible outputs are equiprobable, and   
\begin{equation}
\begin{aligned}
P(r|x,h)&=\Phi\left(\frac{{\rm Re}[y]{\rm Re}[hx]}{\sqrt{1/2}}\right)\Phi\left(\frac{{\rm Im}[r]{\rm Im}[hx]}{\sqrt{1/2}}\right).
\end{aligned}
\end{equation}

\begin{figure}[h]
\begin{center}
\psfrag{C/T}[c][c]{$C/T$ (bits/channel use)}
\psfrag{T}[c][c]{$T$}
{\epsfig{file=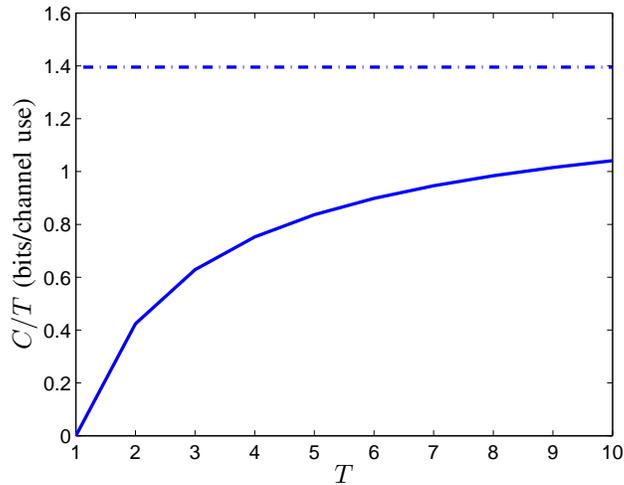, width = 9.2cm}}
\caption{Capacity of the one-bit Rayleigh-fading SISO channel versus coherence interval $T$ for \textrm{SNR}=10dB (dashed curve is for the coherent case $T \rightarrow \infty$).}
\label{CversusT}
\end{center}
\end{figure}

In Fig.~\ref{CversusT}, we plot the normalized capacity of the one-bit Rayleigh-fading SISO channel versus coherence interval $T$ for \textrm{SNR}=10dB. The coherent upper-bound is also plotted.  
\subsubsection{Training Schemes for the 1-bit Block Fading Model}
\begin{figure}[h]
\psfrag{y2}[c][c]{$H \cdot x_{2}$}
  \begin{center}
     \psfrag{/C}[c][c]{}
     \psfrag{T}[c][c]{}
     \psfrag{Training}[c][c]{}
     \psfrag{C}[c][c]{\small $~~~~~~~C^{\rm Training}/C$}
     \psfrag{/T}[c][c]{\small \raisebox{-0.4cm}{$T_{\rm T}/T$}}
     {\epsfig{file=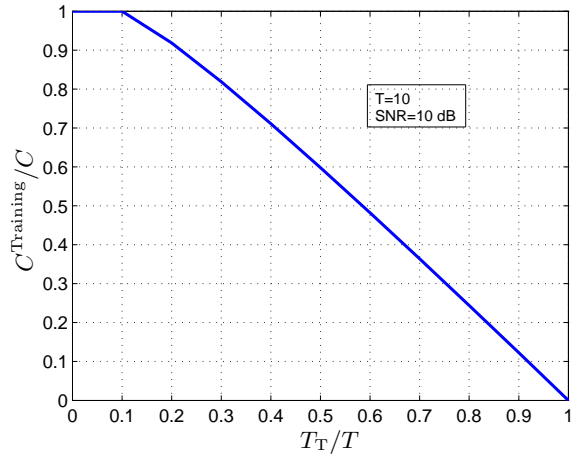, width = 8.5cm}}
     \caption{Achievable rate as a function of the fraction of the coherence time spent on training for $T=10$ and SNR$=10$dB.}
     \label{non-coherent-cap}
  \end{center}
\end{figure}

Training based schemes are attractive for communication over a priori unknown channels, since the receiver task becomes significantly easier. Therefore, most of the wireless communication standards use part of the transmission block for sending a pilot sequence that is known at the receiver:
\begin{equation}
\B{x}= \left[ 
\begin{array}{c}
	\B{x}_{\rm T}  \\
	\B{x}_{\rm D}
\end{array}
 \right],
 \end{equation}
    where $\B{x}_{\rm T}$ is a fixed known training vector of length $T_{\rm T}<T$. This transmit strategy based on the separation of data and training symbols is in general suboptimal from an information theoretical point of view, but is beneficial from a practical point of view.
The achievable rate with joint processing of data and training of this scheme can be written as follows:
\begin{equation}
\begin{aligned}
   C^{\rm Training}_{\rm 1-bit}  &  =  H(\B{r}|\B{x}_{\rm T}) - H(\B{r}|\B{x})                                \\
                     &  =  H(\B{r})-I(\B{x}_{\rm T},\B{r}) - H(\B{r}|\B{x})=C_{{\rm 1-bit},T}-C_{{\rm 1-bit},T_{\rm T}}, 
\end{aligned}
\end{equation}
where we have used the formula for the mutual information $I(\B{x}_{\rm T},\B{r}) =  H(\B{r})-H(\B{r}|\B{x}_{\rm T})=C_{{\rm 1-bit},T_{\rm T}}$. The expression of the capacities with the coherence time can be obtained from (\ref{R_QPSK2}). 
In other words, we have taken the difference of the entropy of the quantized output given the training part $\B{x}_{\rm T}$. In fact, the capacity is reduced by an amount that corresponds to the rate (\ref{R_QPSK2}) with coherence time $T_{\rm T}$:
\begin{equation}
  \begin{aligned}
  C_{{\rm 1-bit},T_{\rm T}} =   2 \sum_{k=0}^{T_{\rm T}}  & \left(\!\!
  \begin{array}{c}
  	T_{\rm T}\\
		k
	\end{array}
	\!\! \right)  \left(\int_{-\infty}^{+\infty}\!\!\!\frac{e^{-\frac{u^2}{2}}}{\sqrt{2\pi}}  \Phi(-\sqrt{\textrm{SNR}}u)^k \Phi(\sqrt{\textrm{SNR}}u)^{T_{\rm T}-k}  {\rm d}u \right) \!\! 	\cdot \\
&\log_2 \left(2^T \int_{-\infty}^{+\infty}\!\!\!\frac{e^{-\frac{u^2}{2}}}{\sqrt{2\pi}}  \Phi(-\sqrt{\textrm{SNR}}u)^k \Phi(\sqrt{\textrm{SNR}}u)^{T_{\rm T}-k}   {\rm d}u  \right).
  \end{aligned}
\end{equation}
 It is worth mentioning that for the case where we fix one symbol in the input sequence $\B{x}$ as a training symbol, we can get the same capacity since $C_{T_{\rm T}=1}=0$. Therefore, in contrast to the unquantized case, a single training symbol for the input sequence $\B{x}$  can always be used without any penalty in the channel capacity. Fig.~\ref{non-coherent-cap} shows the achievable capacity of a SISO channel with a coherence length of $T=10$ and SNR$=10$dB as a function of the training length $T_{\rm T}$. Again, we observe that only one symbol as training will not reduce the capacity, while the curve then decreases with almost a slope of $-1$. Therefore, we can see that choosing the training length to be negligible compared with the coherence time is necessary such that the penalty due to the separation of the channel estimation from the data transmission is small enough. 
  

\subsection{Second-order Expansion of the Mutual Information}
\label{section:mutual_n2}
In this section, we will consider more general channels and elaborate on the second-order expansion of the
input-output mutual information (\ref{transinfo_n}) of the considered system in Fig.~\ref{downlink_figure} as
the signal-to-noise ratio goes to zero, and where only statistical CSI is available at the receiver. We state the main result and we prove it afterwards.  
\newline
\newtheorem {theorem2}[theorem1]{Theorem}
\begin {theorem2}
\label{dualitytheorem_n}
Consider the one-bit quantized MIMO system in Fig.~\ref{downlink_figure}, where the channel matrix $\B{H}$ is zero-mean and circularly distributed under an input distribution $p(\B{x})$  satisfying  $\textrm{E}_{\B{x},\B{H}} [\left\| \B{H} \B{x}\right\|_4^{4+\varepsilon}]<\delta$ for some finite constants $\varepsilon,\delta>0$. Then, to second order, the mutual information (in nats) with statistical CSI between the inputs and the quantized outputs is given by
\begin{equation}
\begin{aligned}	
I(\B{x},\B{r})=& \frac{1}{2} \left(\frac{2}{\pi} \frac{1}{\sigma_\eta^2} \right)^2  {\rm tr} \left\{{\rm E} \left[({\rm nondiag}({\rm E}[\B{H}\B{x}\B{x}^{\rm H}\B{H}^{\rm H}|\B{x}])   )^2 \right]  -   ( {\rm nondiag}({\rm E}[\B{H}{\rm E}[\B{x}\B{x}^{\rm H}]\B{H}^{\rm H}])  )^2  \right\}   \\  
 & + \underbrace{\Delta I(\B{x},\B{r})}_{o(\frac{1}{\sigma_\eta^4})},
\label{transinfo_n_2order}
\end{aligned}
\end{equation}
where $\left\|\B{a} \right\|_4^4$ is the 4-norm of $\B{a}$ taken to the power 4 defined as $\sum_{i,c}a_{i,c}^4$.

\end {theorem2}
\subsubsection{Comments on Theorem \ref{dualitytheorem_n}}
\begin{figure}[h]
\begin{center}
\psfrag{N}[c][c]{$N$}
\psfrag{SNR (linear)}[c][c]{$\frac{P_{\rm Tr}}{\sigma_\eta^2}$}
\psfrag{Mutual Information in bits}[c][c]{\small Mutual Information in bpcu}
{\epsfig {file=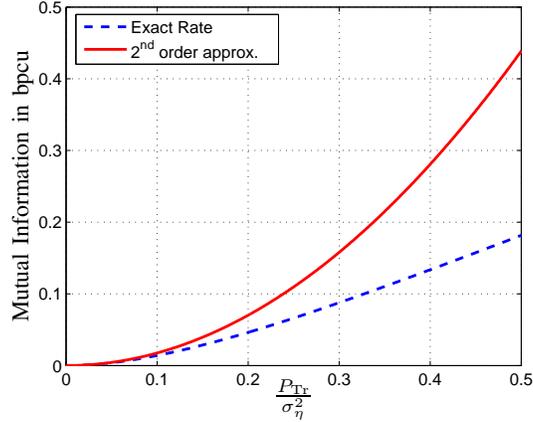, width = 7.8cm}}
\caption{Mutual information of the one-bit block Rayleigh-faded SISO channel with block length 3.}
\label{mutual}
\end{center}
\end{figure}

As an example, Fig.~\ref{mutual} illustrates the rate expression (\ref{rate_T3}) and the quadratic approximation (\ref{transinfo_n_2order}) computed for a  block fading SISO model with a coherence interval of 3 symbol periods ($\B{H}=h\cdot\mathbf{I}_3$, $h \sim \mathcal{CN}(0,1)$)  under QPSK signaling. 

A similar result was derived by Prelov and Verd\`u in \cite{prelov} for the soft output $\B{y}$
\begin{equation}
\begin{aligned}	
I(\B{x},\B{y})=& \frac{1}{2} \frac{1}{\sigma_\eta^4}  {\rm tr} \left\{{\rm E} \left[({\rm E}[\B{H}\B{x}\B{x}^{\rm H}\B{H}^{\rm H} |\B{x}]  )^2 \right]  - ( {\rm E}[\B{H}{\rm E}[\B{x}\B{x}^{\rm H}]\B{H}^{\rm H}] )^2  \right\} 
+o(\frac{1}{\sigma_\eta^4}),
\label{transinfo_2order_unq_n}
\end{aligned}
\end{equation}
 where we identify a power penalty of $\frac{\pi}{2}$ due to quantization and we see that the diagonal elements of ${\rm E}_{\B{H}}[\B{H}\B{x}\B{x}^{\rm H}\B{H}^{\rm H}]$ do not contribute to the mutual information in the hard-decision system. This is because no mutual information can be extracted from the amplitude of each two-level received signal. That means that the channel coefficients have to be correlated, otherwise ${\rm E}_{\B{H}}[\B{H}\B{x}\B{x}^{\rm H}\B{H}^{\rm H}]$ is diagonal and the mutual information is zero. Nevertheless in most systems of practical interest correlation between the channel coefficients, whether temporal or spatial, exists even in multipath rich mobile environments, thus  ${\rm E}_{\B{H}}[\B{H}\B{x}\B{x}^{\rm H}\B{H}^{\rm H}]$ is a rather dense matrix whose Frobenius norm is dominated by the off-diagonal elements rather than the diagonal entries. Thus the low SNR penalty due to the  hard-decision is nearly 1.96 dB for almost all practical channels. 
 This confirms that low-resolution sampling in the low SNR regime performs adequately regardless of the channel model and the kind of CSI available at the receiver, while reducing power consumption. 
\subsubsection{Proof of Theorem \ref{dualitytheorem_n}}
We start again with the definition of the mutual information
\begin{equation}
\begin{aligned}
I(\B{x},\B{r}) & = H(\B{r}) - H(\B{r}|\B{x})                          \\
               & = H({\rm sign}(\B{H}\B{x}+\B{\eta}))  - H({\rm sign}(\B{H}\B{x}+\B{\eta})|\B{x}),
\end{aligned}
\end{equation}

then we use, Lemma~\ref{lemma1_1} and Lemma~\ref{lemma22_2} with $\varepsilon=\frac{1}{\sigma_\eta}$ and $\B{g}(\B{x})=\B{H}\B{x}$ to get the following asymptotic expression:

\begin{equation}
\begin{aligned}
I(\B{x},\B{r})  &=    2N\ln 2 -\frac{2}{\pi}\frac{1}{\sigma_\eta^2} \left\|{\rm E}_{\B{x},\B{H}}[\B{H}\B{x}]\right\|_2^2
 - \frac{1}{\sigma_\eta^4} \left(\frac{2}{\pi^2} {\rm tr} \left(\left({\rm nondiag}({\rm E}_{\B{x},\B{H}}[\B{H}\B{x}\B{x}^{\rm H}\B{H}^{\rm H}]) \right)^2\right) \right.  \nonumber \\
 &~~ \left.-\frac{4}{3\pi}\left\|{\rm E}_{\B{x},\B{H}}[\B{H}\B{x}]\circ {\rm E}_{\B{x},\B{H}}[\B{H}\B{x}\circ\B{H}\B{x}\circ \B{H}\B{x}] \right\|_1^1+\frac{4}{3\pi^2}\left\|{\rm E}_{\B{x},\B{H}}[\B{H}\B{x}]\right\|_4^4\right) \\
 &~~ - 2 N \ln 2 +\frac{2}{\pi} \frac{1}{\sigma_\eta^2} {\rm E}_{\B{x}} [\left\|{\rm E}_{\B{H}}[\B{H}\B{x}]\right\|_2^2]
+ \frac{1}{\sigma_\eta^4} {\rm E}_{\B{x}}  \!\! \left[  \frac{2}{\pi^2} {\rm tr}\left(\left({\rm nondiag}({\rm E}_{\B{H}}[\B{H}\B{x}\B{x}^{\rm H}\B{H}^{\rm H}]) \right)^2\right) \right.  \nonumber \\
 &~~\left.-\frac{4}{3\pi}\left\|{\rm E}_{\B{H}}[\B{H}\B{x}]\circ {\rm E}_{\B{H}}[\B{H}\B{x} \circ \B{H}\B{x} \circ \B{H}\B{x}] \right\|_1^1+ \frac{4}{3\pi^2}\left\|{\rm E}_{\B{H}}[\B{H}\B{x}]\right\|_4^4\right] +o(\frac{1}{\sigma_\eta^4})  \\
 &= \frac{2}{\sigma_\eta^4 \pi^2} {\rm tr} \left( {\rm E}_{\B{x}} [ \left( {\rm nondiag}({\rm E}_{\B{H}}[\B{H} \B{x}  \B{x}^{\rm H} \B{H}^{\rm H}])\right)^2 ] -   \left( {\rm nondiag}({\rm E}_{\B{x},\B{H}}[\B{H} \B{x}  \B{x}^{\rm H} \B{H}^{\rm H}])\right)^2  \right) + o(\frac{1}{\sigma_\eta^4}),
\end{aligned}
\end{equation}
where the last step follows from the fact that the channel matrix $\B{H}$ has zero-mean. We note that since $\B{H}$ has a proper distribution, then  $\B{g}(\B{x})=\B{H}\B{x}$ also has a zero-mean proper distribution, fulfilling $p(\B{H})=p({\rm j}\B{H x})$. Therefore the conditions of Theorem~\ref{low_snr_mi} are fulfilled. 
   This yields the result stated by the theorem and completes the proof. 
Note that the condition $\textrm{E}_{\B{x},\B{H}} [\left\|\B{H}(\B{x})\right\|_4^{4+\varepsilon}] < \delta$ for some finite constants $\varepsilon,\delta>0$ is necessary, so that the remainder of the expansion is asymptotically negligible as shown in the coherent case (see Section~\ref{section:duality}). 


\subsection{IID Block Rayleigh Fading MIMO Channels}
We consider as an example a  point-to-point quantized MIMO channel where the transmitter employs $M'$ antennas and the  receiver has $N'$ antennas. We assume a block-Rayleigh fading model \cite{marzetta}, in which the channel propagation matrix $\B{H}' \in \mathbb{C}^{M'\times N'}$ remains constant for a coherence interval of length $T$ symbols, and then changes to a new independent value. The entries of the channel matrix are i.i.d. zero-mean complex circular Gaussian with unit variance. The channel realizations are unknown to both the transmitter and receiver. At each coherence interval, a sequence of vectors $\B{x}_1,\ldots,\B{x}_T$ are transmitted at each time slot via the multiple antennas. The transmitted and received signal matrices are then related as follows (total dimensions: $N=TN'$ and $M=TM'$)
 
\begin{equation}
\underbrace{    \left[ 
\begin{array}{c}
\B{y}_1 \\
\B{y}_2 \\
\vdots \\
\B{y}_T
\end{array}
\right]}_{\B{y}} = 
\underbrace{\left[ 
\begin{array}{cccc}
\B{H}' & \B{0} & \cdots & \B{0} \\
\B{0} & \B{H}' & \cdots & \B{0} \\
\B{0} & \B{0} & \ddots & \B{0} \\
\B{0} & \B{0} & \cdots & \B{H}' 
\end{array}
\right]}_{\B{H}} \cdot 
  \underbrace{  \left[ 
\begin{array}{c}
\B{x}_1 \\
\B{x}_2 \\
\vdots \\
\B{x}_T
\end{array}
\right]}_{\B{x}} + 
\underbrace{\left[ 
 \begin{array}{c}
 \B{\eta}_1 \\
 \B{\eta}_2 \\
 \vdots \\
 \B{\eta}_T 
\end{array}
\right]}_{\B{\eta}}.
\end{equation}
Then the expected value of the received signal conditioned on the input is given by  
\begin{equation}
{\rm E}[\B{H}\B{x}\B{x}^{\rm H}\B{H}^{\rm H} |\B{x}]  = \left[
\begin{array}{cccc}
{\rm E} [ \B{H}'\B{x}_1 \B{x}_1^{\rm H} \B{H}'^{\rm H}]  & {\rm E} [ \B{H}'\B{x}_1 \B{x}_2^{\rm H} \B{H}'^{\rm H}] & \cdots & {\rm E} [ \B{H}'\B{x}_1 \B{x}_T^{\rm H} \B{H}'^{\rm H}] \\
{\rm E} [ \B{H}'\B{x}_2 \B{x}_1^{\rm H} \B{H}'^{\rm H}] & {\rm E} [ \B{H}'\B{x}_2 \B{x}_2^{\rm H} \B{H}'^{\rm H}] & \cdots & {\rm E} [ \B{H}'\B{x}_2 \B{x}_T^{\rm H} \B{H}'^{\rm H}] \\
 \ddots & \ddots & \ddots & \ddots \\
{\rm E} [ \B{H}'\B{x}_T \B{x}_1^{\rm H} \B{H}'^{\rm H}] & {\rm E} [ \B{H}'\B{x}_T \B{x}_2^{\rm H} \B{H}'^{\rm H}] & \cdots &{\rm E} [ \B{H}'\B{x}_T \B{x}_T^{\rm H} \B{H}'^{\rm H}] \\
\end{array}
   \right].
\end{equation}
Since $\B{H}'$ is i.i.d. distributed, it can be shown that ${\rm E}_{\B{H}'} [ \B{H}'\B{x}_i \B{x}_j^{\rm H} \B{H}'^{\rm H}] =   \B{x}_i^{\rm T} \B{x}_j^* \cdot \textbf{I}_{N'} $. Therefore, we have 
\begin{equation}
{\rm E}[\B{H}\B{x}\B{x}^{\rm H}\B{H}^{\rm H} |\B{x}]  = ( \B{X} \cdot \B{X}^{\rm H} ) \otimes  \textbf{I}_{N'},
\end{equation}
where the rows of the matrix $\B{X} \in \mathbb{C}^{T \times M'}$ are the vectors $\B{x}_i^{\rm T}$, for $1\leq i \leq T$. With this,  (\ref{transinfo_n_2order})  asymptotically becomes
\begin{equation}
\begin{aligned}	
I(\B{x},\B{r})=& \frac{N'}{2} \left(\frac{2}{\pi} \frac{1}{\sigma_\eta^2} \right)^2  {\rm tr} \left\{{\rm E} \left[({\rm nondiag}(  \B{X} \cdot \B{X}^{\rm H}  ))^2 \right]  -   ( {\rm nondiag}({\rm E} [\B{X} \cdot \B{X}^{\rm H}] ))^2  \right\}  + o(\frac{1}{\sigma_\eta^4}),
\end{aligned}
\end{equation}
while for the ideal case, we get from (\ref{transinfo_2order_unq_n})  (see also \cite{rao})
\begin{equation}
\begin{aligned}	
I(\B{x},\B{y})=& \frac{N'}{2} \left(\frac{1}{\sigma_\eta^2} \right)^2  {\rm tr} \left\{{\rm E} \left[(  \B{X} \cdot \B{X}^{\rm H}  )^2 \right]  -    ({\rm E} [\B{X} \cdot \B{X}^{\rm H}] )^2  \right\}  + o(\frac{1}{\sigma_\eta^4}).
\end{aligned}
\end{equation}
Now, assuming i.i.d. Gaussian inputs with  $\mathrm {E}[\B{X}\B{X}^\textrm{H}]=  P_{\rm Tr} \cdot \textbf{I}_{T}$, then we obtain finally (c.f. expectation (\ref{EH2}))
\begin{equation}
\begin{aligned}	
I(\B{x},\B{r})=& \frac{N'}{2} \left(\frac{2}{\pi} \frac{P_{\rm Tr}}{\sigma_\eta^2} \right)^2 T(T-1) + o(\frac{1}{\sigma_\eta^4})\\
I(\B{x},\B{y})=& \frac{N'}{2} \left(\frac{P_{\rm Tr}}{\sigma_\eta^2} \right)^2 T^2 + o(\frac{1}{\sigma_\eta^4}).
\end{aligned}
\end{equation}
Evidently, in order for $I(\B{x},\B{r})$ and $I(\B{x},\B{y})$ to be equal with the same power, we need to increase the number of one-bit receive antennas $N'$ by roughly a factor of $\pi^2/4$ when the coherence interval satisfies $T \gg 1$. The same result has been obtained in \cite{YongzhiGlobecom} with a pilot-based scheme and using the pseudo-quantization noise model.  
\subsection{SIMO Channels with Delay Spread and Receive Correlation at low SNR}
\label{section:app_simo}
As a further example, we use the result from  Theorem \ref{dualitytheorem_n} to compute the low SNR mutual information of a frequency-selective single input multi-output (SIMO) channel with delay spread and receive  correlation  both in time and space and obtain the asymptotic achievable rate  under average and peak power constraints. The quantized output of the considered model at time $k$ is 
\begin{equation} 
\begin{aligned}
\B{r}_k &= Q(\B{y}_k)  \\
\B{y}_k &=\sum_{t=0}^{T-1} \B{h}_k[t] x_{k-t}  +\B{\eta}_k \in \mathbb{C}^{N'},
\end{aligned} 
\label{convhx}
\vspace{-0.1cm}
\end{equation}
where the noise process $\{\B{\eta}_k\}$ is i.i.d. in time and space, while the $T$ fading processes $\{\B{h}_k[t]\}$ at each tap $t$ are assumed to be independent zero-mean proper complex Gaussian processes. Furthermore, we assume a separable temporal spatial correlation model, i.e. 
\begin{equation}
{\rm E}[\B{h}_k[t] \B{h}_{k'}[t']^{\rm H}]= \B{C}_h \cdot c_h(k-k') \alpha_t \delta[t-t'].
\end{equation}
Here, $\B{C}_h$ denotes the receive correlation matrix, $c_h(k)$ is the autocorrelation function of the fading process, and the scalars $\alpha_i$ represent  the power-delay profile. The correlation parameters can be normalized so that
\vspace{-0.1cm} 
\begin{equation}
{\rm tr}(\B{C}_h)=N',~ c_h(0)=1 \textrm{ and } \sum_{t=0}^{T-1} \alpha_t=1.
\label{normalization}
\vspace{-0.1cm} 
\end{equation} 
In other words, the energy
in  each receive antenna's impulse response equals one on
average. 
On the other hand we assume that the transmit signal $x_k$ is subject to an average power constraint ${\rm E}[\| x_k\|^2]\leq P_{\rm Tr}$ and a peak power constraint $|x_k|^2 \leq \beta \cdot P_{\rm Tr}$, $\forall k$, with $\beta \geq 1$. It should be pointed out that a peak power constraint constitutes a stronger condition than necessary for the validity of Theorem~\ref{dualitytheorem_n}, involving just a fourth-order moment constraint on the input.  
We consider now a time interval of length $n$ (a block of $n$ transmissions). Collect a vector sequence $\B{y}_k$ of length $n$ into the vector $\B{y}$ as
\begin{equation}
\B{y}^{\!(\!n\!)}=\left[\B{y}^\textrm{T}_{n-1},\ldots,\B{y}^\textrm{T}_0 \right]^\textrm{T},
\end{equation}
and form the block cyclic-shifted matrix $\B{H}^{\!(\!n\!)}\in \mathbb{C}^{(N' n)\times n}$
\begin{equation}
\B{H}^{\!(\!n\!)}\!=\!\left[
\begin{array}{ccccc}
\!\!\B{h}_{n-1}[0]\!\!\!& \cdots &\!\! \B{h}_{n-1}[T\!-\!1]\!\!&0&\cdots\\
&\ddots& &&\ddots\\
\ddots	&&  &\ddots&\\
	\cdots &\!\!\B{h}_{0}[T\!-\!1]\!\!\!&0~~~~~\cdots&0& \!\!\B{h}_{0}[0] \!\!
\end{array}
\right].
\end{equation}
With 
\begin{equation}
\B{x}^{\!(\!n\!)}=\left[x_{n-1},\ldots,x_{0}\right]^\textrm{T},
\end{equation}
and $\B{\eta}^{\!(\!n\!)}$ and $\B{r}^{\!(\!n\!)}$ defined similar to $\B{y}^{\!(\!n\!)}$, the following quantized space-time model may be formulated as a (loose) approximation of (\ref{convhx}), where the resulting  dimensions are $N=N' n$ and $M=n$
\vspace{-0.1cm} 
\begin{equation}
 \B{r}^{\!(\!n\!)}=Q( \B{y}^{\!(\!n\!)})= Q\left(\B{H}^{\!(\!n\!)}\B{x}^{\!(\!n\!)}+\B{\eta}^{\!(\!n\!)} \right).
\vspace{-0.1cm} 
\end{equation}
\subsubsection{Asymptotic Achievable Rate}
Now, we elaborate on the asymptotic information rate of this channel setting. First we establish an upper bound and then we examine its achievability.
\newtheorem {proposition1}{Proposition}
\begin {proposition1}
If $c_h(k)$ is square-summable, then the mutual information of the described space-time model admits the following asymptotic upper bound, for any distribution fulfilling the average and peak power constraints
\vspace{-0.1cm} 
\begin{equation} 
\begin{aligned}
\lim_{n \rightarrow \infty }\frac{1}{n}I(\B{x}^{\!(\!n\!)},\B{r}^{\!(\!n\!)}) \leq  \left(\frac{2}{\pi} \frac{P_{\rm Tr}}{\sigma_\eta^2} \right)^2  U(\beta),
 \end{aligned}
 \label{upper}
\end{equation}
where
\begin{equation} 
 \begin{aligned}
 U(\beta)=\left\{
\begin{array}{ll}
\beta\cdot \zeta +(\beta-1)\chi & \textrm{for } \beta(\zeta+\chi)\geq 2 \chi \\
\beta^2\frac{(\zeta+\chi)^2}{4\chi} & \textrm{else, } 
\end{array}
 \right.
 \end{aligned}
\end{equation}
\begin{equation} 
 \begin{aligned}
 \zeta={\rm tr}(\B{C}_h^2)\sum_{k=1}^{\infty}c_h(k)^2
 \end{aligned}
 \label{lambda}
\end{equation}
and
\begin{equation} 
 \begin{aligned}
 \chi=\frac{1}{2}{\rm tr}(({\rm nondiag}(\B{C}_h))^2) c_h(0).
 \end{aligned}
 \label{sigma}
\end{equation}
\label{proposition1}
\end {proposition1}
\begin{proof} 
Due to the peak power constraints, the conditions of Theorem~\ref{dualitytheorem_n} are satisfied; thus the second order approximation (\ref{transinfo_n_2order}) is valid. A tight upper bound is obtained by looking at the maximal value that can be achieved by  the expression (\ref{transinfo_n_2order}) up to second order. We do this  in two steps. We  first maximize the trace expression in (\ref{transinfo_n_2order}) under a prescribed average power per symbol $\gamma$. The maximum can be, in turn, upper bounded by the supremum of the first term minus the infimum of the second term under the prescribed average power and the original peak power constraint. After that, we perform an optimization over the parameter $\gamma$ itself. That is 
\begin{equation}
\begin{aligned}	
I(\B{x}^{\!(\!n\!)},\B{r}^{\!(\!n\!)})\leq \frac{1}{2} \left(\frac{2}{\pi} \frac{1}{\sigma_\eta^2} \right)^2  \max_{0\leq\gamma\leq 1}   \Bigg\{ &  \sup_{ \stackrel {|x_k|^2\leq \beta P_{\rm Tr}, \forall k}{{\rm E}[\|\B{x}^{\!(\!n\!)}\|^2]= \gamma P_{\rm Tr} \cdot  n}
 } \!\!\!\! \!\!\!\!{\rm tr}\! \left(\!{\rm E}\! \left[\!({\rm nondiag}({\rm E}[\B{H}^{\!(\!n\!)}\B{x}^{\!(\!n\!)}\B{x}^{\!(\!n\!),{\rm H}}\B{H}^{\!(\!n\!),{\rm H}}|\B{x}])   )^2 \right] \right) \\
&\!\!\! -\!\!\!\!\!\!\!\! \inf_{ \stackrel {|x_k|^2\leq \beta P_{\rm Tr}, \forall k}{{\rm E}[\|\B{x}^{\!(\!n\!)}\|^2\!]= \gamma P_{\rm Tr} \cdot n}}\!\! \!\!\!\!\!\!{\rm tr} \!\left(\!( {\rm nondiag}({\rm E}[\B{H}^{\!(\!n\!)}{\rm E}\![\B{x}^{\!(\!n\!)}\B{x}^{\!(\!n\!),{\rm H}}]\B{H}^{\!(\!n\!),{\rm H}}])  )^2 \!  \right) \!\! \!\Bigg\}
\!\!+\!\!{o(\frac{1}{\sigma_\eta^4})}.
\label{transinfo_upper}
\end{aligned}
\end{equation}
Evaluating the expectation with respect to the channel realizations, we get
\begin{equation}
 {\rm E}[\B{H}^{\!(\!n\!)}\B{x}^{\!(\!n\!)}\B{x}^{\!(\!n\!),{\rm H}}\B{H}^{\!(\!n\!),{\rm H}}] = \B{D} \otimes \B{C}_h,
\end{equation}
with
\begin{equation}
d_{k,k'}= c_h(k-k') \sum_{t=0}^{T-1} \alpha_t x_{k-t} x_{k'-t}^* \textrm{ for } k,k' \in \{ 0,\ldots,n-1 \}.
\end{equation}
Therefore
\begin{equation}
\begin{aligned}	
  {\rm tr}\! \left(\!({\rm nondiag}({\rm E}[\B{H}^{\!(\!n\!)}\B{x}^{\!(\!n\!)}\B{x}^{\!(\!n\!),{\rm H}}\B{H}^{\!(\!n\!),{\rm H}}|\B{x}])   )^2  \right) = &  \sum_{k=0}^{n-1} \sum_{\stackrel{k'=0}{k' \neq k}}^{n-1}  {\rm tr}(\B{C}_h^2) c_h(k-k') \left| \sum_{t=0}^{T-1}  \alpha_t x_{k-t}  x_{k'-t}^* \right|^2 + \\
  & \sum_{k=0}^{n-1}  {\rm tr}(({\rm nondiag}(\B{C}_h))^2) c_h(0) \left| \sum_{t=0}^{T-1}  \alpha_t |x_{k-t}|^2  \right|^2.
\end{aligned}	
\end{equation}
It turns out that the supremum of the first trace term is achieved when all $x_k$ inputs take,  simultaneously during the considered time interval, either the value zero, or the peak value $\beta$ with a duty cycle of $\gamma\beta^{-1}$. On the other hand, the infimum of the second trace expression is obtained, under the prescribed average power, when ${\rm E}[\B{x}\B{x}^{\rm H}]=\gamma P_{\rm Tr} {\bf I}_n$. Calculation shows that
\begin{equation}
\begin{aligned}	
\!\!\frac{1}{n}I(\!\B{x}^{\!(\!n\!)},\B{r}^{\!(\!n\!)}\!)\!\! \leq &\!\! \max_{0\leq  \gamma \leq 1} \!\! \left(\!\frac{2}{\pi} \frac{P_{\rm Tr}}{\sigma_\eta^2} \!\right)^{\!\!2}\!\!\left(\sum_{t=0}^{T-1}\! \alpha_t \!\!\right)^{\!\!\!2} \!\! \Bigg[\!\gamma\beta \!\cdot\!\underbrace{{\rm tr}(\!\B{C}_h^2)\!\!\sum_{k=1}^{n-1}(1\!-\!\frac{k}{n})c_h(k)^2}_{\zeta^{\!(\!n\!)}}  \\
&\!\!+  \gamma(\beta-\gamma) \underbrace{\frac{1}{2}{\rm tr}(({\rm nondiag}(\B{C}_h))^2) c_h(0) }_{\chi}\Bigg] +o(\frac{1}{\sigma_\eta^4}).
\end{aligned}
\end{equation}
Now, the maximization over $\gamma$ delivers 
\begin{equation}
\begin{aligned}	
\gamma_{\rm opt}= \min \{1,\beta \frac{\zeta^{\!(\!n\!)}+\chi}{2\chi}\}.
\end{aligned}
\label{gamma_opt}
\end{equation}
Thus, taking the limit $n \rightarrow \infty$ yields $\zeta^{\!(\!n\!)} \rightarrow \zeta$ as in  (\ref{lambda}) and by the normalizations in (\ref{normalization}), we end up with the result of the proposition.\footnote{Observe that $\zeta$ and $\chi$ are indicators for the temporal and spatial coherence, respectively.}
\end{proof}
We next turn to the question of whether the upper bound  suggested by Proposition \ref{proposition1} is achievable. A closer examination of (\ref{transinfo_upper}) demonstrates that the upper bound could be achieved if the input distribution satisfies the following condition, for any time instants $k$ and $l$ within the block of length $n$:
\begin{equation}
\begin{aligned}	
x_k x_l^*= a {\rm e}^{{\rm j}\Omega(k-l)},
\end{aligned}
\end{equation}
for some $\Omega\neq 0$ and random $a\in\{0,\sqrt{\beta \cdot P_{\rm Tr}} \}$, while having ${\rm E}[\B{x}^{\!(\!n\!)}\B{x}^{\!(\!n\!),{\rm H}}]=\gamma P_{\rm Tr} {\bf I}_{n}$ as already mentioned in the proof of Proposition \ref{proposition1}. Clearly an \emph{on-off frequency-shift keying} (OOFSK) modulation for the input, as follows,  can fulfill these conditions  
\begin{equation}
\begin{aligned}
x_k=Z\cdot {\rm e}^{{\rm j}k\Omega},~~k\in\{1,\ldots,n\},	 
\end{aligned}
\end{equation}
where $Z$ takes the value $\sqrt{\beta P_{\rm Tr}}$ with probability $\gamma_{\rm opt} \beta^{-1}$ and zero with probability $(1-\gamma_{\rm opt} \beta^{-1})$, and $\Omega$ is uniformly distributed over the set $\{\frac{2\pi}{n} ,\ldots,\frac{2\pi(n-1)}{n} \}$. This is similar to the results  of  the unquantized case \cite{sethuraman,telatar}. 
\subsubsection{Discussion}
\begin{figure}[h]
\begin{center}
\psfrag{T=1}[c][c]{\footnotesize $T=1$}
\psfrag{T=5}[c][c]{\footnotesize $T=5$}
\psfrag{sigma/mu}[c][c]{$\frac{\chi}{\zeta}$}
{\epsfig {file=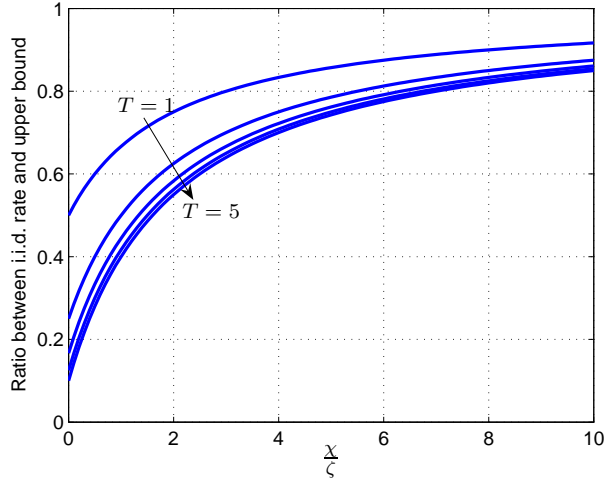, width = 9cm}}
\caption{Ratio of i.i.d. rate (\ref{iidrate}) and upper bound (\ref{upper}) vs. the spatial-to-temporal coherence ratio $\frac{\chi}{\zeta}$ for $\beta=2$ and uniform delay spread, i.e.  $\alpha_t=\frac{1}{T}$.}
\label{iid_rate_figure}
\end{center}
\vspace{-0.5cm}
\end{figure}

We notice from  (\ref{gamma_opt}) that the average power constraint ${\rm E}[\| x_k\|^2]\leq P_{\rm Tr}$ is only active when $\beta(\zeta+\chi) \geq 1$, which means that it is not necessarily optimal to utilize the total available or allowed average power especially if a tight peak power constraint is present. In addition, if we impose only a peak power constraint, i.e. $\beta=1$, then if $\zeta< \chi$ the on-off strategy with the zero symbol ($a=0$) is required to approach the capacity. We notice that on-off modulation in this case is crucial and should not be regarded as duty cycle transmission where the link remains inactive for a certain amount of time. 
We observe also from Proposition \ref{proposition1} that spreading the power over different taps does not affect the low SNR rate, while receiver correlation is beneficial due to two effects. First the mutual information increases with $\chi$  defined in (\ref{sigma}) which is related to the norm of the off-diagonal elements of $\B{C}_h$. Second, under the normalization ${\rm tr}(\B{C}_h)=N'$, the Frobenius norm ${\rm tr}(\B{C}_h^2)$ increases with more correlation among the receive antennas, and consequently, higher rates at low SNR can be achieved due to relation (\ref{lambda}). In fact, both spatial and temporal correlations are extremely beneficial at low SNR, even more than in the unquantized case.
Besides we note that the achievability of the upper bound stated in Proposition \ref{proposition1}, as discussed previously, is obtained at the cost of burstiness  in frequency which may not agree with some 
specifications imposed on many systems.\footnote{Note that such observations hold also in the unquantized case \cite{sethuraman}.} Therefore, it is interesting to look at the asymptotic rate of i.i.d. input symbols drawn from the set
$\{ -\sqrt{\beta P_{\rm Tr}},0, \sqrt{\beta P_{\rm Tr} } \}$. In that case it turns out by (\ref{transinfo_n_2order}) that
  \begin{equation}
\begin{aligned}
\frac{1}{n}I_{\rm IID}(\B{x}^{\!(\!n\!)},\B{r}^{\!(\!n\!)}) \!\!\approx \!\!& \max_{0\leq  \gamma \leq 1} \left(\!\frac{2}{\pi} \frac{P_{\rm Tr}}{\sigma_\eta^2} \!\right)^{\!\!2}\!\! \left[ \gamma^2\zeta^{\!(\!n\!)} \sum_{t=0}^{T-1} \alpha_t^2+  \gamma(\beta-\gamma) \chi \left( \sum_{t=0}^{T-1} \alpha_t \right)^2 \right]\!\!.
\end{aligned}
\label{iidrate}
\end{equation}
Here, we observe that, contrary to to the FSK-like scheme, the mutual information with i.i.d input is negatively affected by the delay spread since $\sum_{t=0}^{T-1} \alpha_t^2\leq 1$ by the normalization (\ref{normalization}). Nevertheless for the case $\chi \gg \zeta$, i.e. low temporal correlation (high Doppler spread), the gap to the upper bound in Proposition \ref{proposition1}  vanishes, as demonstrated in Fig.~\ref{iid_rate_figure}.
\section{Conclusion}
\label{section:conclusion}
Motivated by the simplicity of one-bit ADCs for sampling large dimensional signals, we provided a general second order asymptotic analysis for the entropy of one-bit quantized vector signals. We used these results to evaluate the mutual information around zero SNR for a wide class of channel models.  We have shown that the reduction of low SNR channel capacity by a factor of $2/\pi$ due to 1-bit quantization holds for the general MIMO case with uncorrelated noise. Additionally, the  non-coherent MIMO channel was studied in detail and a similar conclusion was made. Thereby, as one might expect, it turns out that QPSK plays an essential role in achieving optimal performance for 1-bit quantized channels at low SNR.  

\appendices
\section{Proof of Theorem~\ref{low_snr_mi}}
\label{proof_theorem1}

We aim at the derivation of the second order approximation (Taylor expansion) of the entropy $H(\B{r})$, where $\B{r}={\rm sign}(\B{y})={\rm sign}(\varepsilon \B{x}+ \B{\eta})$, with respect to $\varepsilon$. Thereby, we assume that the vector $\B{\eta}$ is i.i.d. Gaussian distributed with unit variance and $\B{x}$ has a general distribution. As will be shown in the following,  this second order Taylor expansion is a function of the moments of the distribution of $\B{x}$, which have to fulfill certain conditions. 
For simplicity,  we consider first the real-valued case for the computation of the second order approximation of the entropy  $H({\rm sign}(\varepsilon \B{x}+ \B{\eta}))$ assuming that $\B{x}$ has a real valued distribution and $\B{\eta}$ follows a real-valued Gaussian distribution with variance $\frac{1}{2}$. The generalization to the complex case can be obtained based on the real-valued representation of the complex channel and is performed at the end of the proof. 
 The probability mass function $P_\epsilon(\B{r})$ of the random vector $\B{r}={\rm sign}(\varepsilon \B{x}+ \B{\eta}) \in\{ \pm 1\}^N$  is given by 
\begin{equation}
P_\epsilon(\B{r})= P(\B{r}\circ \B{y} \in \mathbb{R}_+^N) ={\rm E}_{\B{x}} \int_{\mathbb{R}_+^N} p_\epsilon ( \B{r} \circ \B{y}| \B{x} ) {\rm d}\B{y},
\end{equation}
where
\begin{equation}
   p_\epsilon(\B{y}|\B{x})=\frac{1}{\pi^{N/2}}{\rm e}^{-\left\|\B{y}-\varepsilon \B{x}  \right\|_2^2}.
\end{equation}
We also have the general expression of the entropy (in nats) of random vector $\B{r}$ as
\begin{equation}
  H(\B{r})= \sum_{\B{r}\in\{\pm 1\}^{N}} -P_\varepsilon (\B{r}) \ln P_\varepsilon (\B{r}).
\end{equation}  

Next, based on the derivatives of the function $-z\ln z$, we calculate the second order expansion of the entropy with respect to $\varepsilon$. For that, we first notice that the linear and cubic terms of the Taylor expansion vanish due to the fact that the entropy function $H({\rm sign}(\varepsilon \B{x} +\B{\eta} ))$ is an even function with respect to  $\varepsilon$ since $H({\rm sign}(\varepsilon \B{x}+ \B{\eta}))=H({\rm sign}(-\varepsilon \B{x}+ \B{\eta}))$. Additionally, since $P_\varepsilon(\B{r})$ is a probability mass function, i.e.,  $\sum_{\B{r}} P_\varepsilon(\B{r}) = 1$, we have
\begin{equation}
   \sum_{\B{r}} P_\varepsilon^{(k)}(\B{r}) = 0, 
\end{equation}
for any $k$-th derivative of $P_\varepsilon^{(k)}(\B{r})$ with respect to $\varepsilon$.   
Based on these observations, we get the following second order expansion
\begin{equation}
\begin{aligned}
H(\B{r}) & =
\sum_{\B{r}}-P_0(\B{r} )\ln P_0(\B{r}) -\varepsilon ^2 \frac{P_0'(\B{r} )^2}{2 P_0(\B{r} ) }-   \\
&~~~~~ \varepsilon ^4\left( \frac{P_0'(\B{r})^4}{12P_0(\B{r})^3 }- \frac{P_0'(\B{r})^2P_0''(\B{r})}{4P_0(\B{r})^2}+   \frac{P_0''(\B{r})^2}{8 P_0(\B{r}) }+ \frac{P_0'(\B{r})P_0^{(3)}(\B{r})}{6P_0(\B{r})} \right)  + \Delta H(\B{r}),
\end{aligned}
\label{entropy_order_2}
\end{equation}
where $P_0'(\B{r})$, $P_0''(\B{r})$ and  $P_0'(\B{r})P_0^{(3)}(\B{r})$  are the first, second and third order derivatives of $P_\varepsilon(\B{r})$ at $\varepsilon=0$, which will be derived in the following. To this end, we utilize the following derivatives of the unquantized output distribution $p_\epsilon(\B{y}|\B{x})=\frac{1}{\pi^{N/2}}{\rm e}^{-\left\|\B{y}-\varepsilon \B{x}  \right\|_2^2}$
\begin{equation}
\begin{aligned}
     p_0(\B{y}|\B{x})= \frac{1}{\pi^{N/2}}{\rm e}^{-\left\|\B{y} \right\|_2^2},
\end{aligned}
\end{equation}
\begin{equation}
\begin{aligned}
     p_0'(\B{y}|\B{x})= \frac{1}{\pi^{N/2}}{\rm e}^{-\left\|\B{y} \right\|_2^2} 2 \B{y}^{\rm T}  \B{x},
\end{aligned}
\end{equation}
\begin{equation}
\begin{aligned}
     p_0''(\B{y}|\B{x})= \frac{2}{\pi^{N/2}}{\rm e}^{-\left\|\B{y} \right\|_2^2} ( 2 (\B{y}^{\rm T} \B{x})^2 -\left\|\B{x}\right\|_2^2 ),
\end{aligned}
\end{equation}
\begin{equation}
\begin{aligned}
     p_0^{(3)}(\B{y}|\B{x})  = \frac{6}{\pi^{N/2}}{\rm e}^{-\left\|\B{y} \right\|_2^2} 2\B{y}^{\rm T} \B{x} ( \frac{2}{3} (\B{y}^{\rm T} \B{x} )^2 -\left\|\B{x}\right\|_2^2 ).
\end{aligned}
\end{equation}

Now, we calculate $P_0(\B{r})$ 
\begin{align}
P_0(\B{r})=  \int_{\mathbb{R}_+^N}  \frac{2}{\pi^\frac{N}{2}}{\rm e}^{-\left\|\B{y}\right\|_2^2} {\rm d}\B{y} = \frac{1}{2^N}. 
\end{align}
Afterward, we evaluate the first order derivative as  
\begin{align}
P_0'(\B{r}) = \int_{\mathbb{R}_+^N}  \frac{2}{\pi^\frac{N}{2}}{\rm e}^{-\left\|\B{y}\right\|_2^2}(\B{r}\circ\B{y})^{\rm T} {\rm E} [\B{x}] {\rm d}\B{y}= \frac{1}{2^N} \frac{2}{\sqrt{\pi}}  \B{r}^{\rm T} {\rm E} [\B{x}],
\end{align}
while the second order derivative reads as

\begin{equation}
\begin{aligned}
P_0''(\B{r}) &= {\rm E}_{\B{x}} \int_{\mathbb{R}_+^N}  \frac{2}{\pi^{N/2}}{\rm e}^{-\left\|\B{y}\right\|_2^2}  \left( 2 ((\B{r}\circ\B{y})^{\rm H} \B{x} )^2 -\left\|\B{x}\right\|_2^2 \right)  {\rm d}\B{y} \\
&={\rm E}_{\B{x}} \int_{\mathbb{R}_+^N}  \frac{2}{\pi^{N/2}}{\rm e}^{-\left\|\B{y}\right\|_2^2}  \left(  \B{x}^{\rm T} \left( 2 (\B{r}\circ\B{y})(\B{r}\circ\B{y})^{\rm T}  - 1 \right) \B{x}    \right)  {\rm d}\B{y}\\
&= \frac{1}{2^N} \frac{4}{\pi}  {\rm E} \left[ \B{x}^{\rm T} {\rm nondiag} \left(  \B{r}\B{r}^{\rm T} \right) \B{x} \right]  \\
&= \frac{1}{2^N} \frac{4}{\pi} \B{r}^{\rm T} {\rm nondiag} \left({\rm E} [\B{x}\B{x}^{\rm T}]\right) \B{r},
\end{aligned}
\end{equation}
where we used the following result
\begin{equation}
\begin{aligned}
\int_{\mathbb{R}_+^N}   \frac{{\rm e}^{-\| \B{y}\|^2}}{\pi^{N/2}} (y_i \cdot r_i)(y_j \cdot  r_j) {\rm d}\B{y}=\left\{
\begin{array}{ll}
\frac{1}{2\cdot 2^N}	 & \textrm{for } i=j  \\
	\frac{r_ir_j}{\pi \cdot  2^N}	 & {\rm else,}
\end{array}
  \right. 
 \end{aligned}
\end{equation}
and the relation ${\rm tr}({\rm nondiag}(\B{A})\B{B})= {\rm tr}(\B{A}{\rm nondiag}(\B{B}))$ for any two quadratic matrices $\B{A}$ and $\B{B}$.

In a similar way, we calculate  $P_0^{(3)}(\B{r} )$
\begin{equation}
\begin{aligned}
P_0^{(3)}(\B{r} )&= {\rm E}_{\B{x}} \int_{\mathbb{R}_+^N}  \frac{6}{\pi^N}{\rm e}^{-\left\|\B{y}\right\|_2^2}2 ((\B{r}\circ\B{y})^{\rm T} \B{x} )  \left( \frac{2}{3} ((\B{r}\circ\B{y})^{\rm T} \B{x} )^2 -\left\|\B{x}\right\|_2^2 \right)  {\rm d}\B{y} \\
&={\rm E}_{\B{x}} \int_{\mathbb{R}_+^N}  \left( \frac{8}{\pi^N}{\rm e}^{-\left\|\B{y}\right\|_2^2}   ((\B{r}\circ\B{y})^{\rm T} \B{x})^3 -  \frac{12}{\pi^N}{\rm e}^{-\left\|\B{r}\right\|_2^2}   ((\B{r}\circ\B{y})^{\rm T} \B{x}) \left\|\B{x}\right\|_2^2  \right)  {\rm d}\B{y} \\
&=  {\rm E}_{\B{x}} \int_{\mathbb{R}_+^N}  \frac{8}{\pi^N}{\rm e}^{-\left\|\B{y}\right\|_2^2}   ((\B{r}\circ\B{y})^{\rm T} \B{x})^3  {\rm d}\B{y}     - \frac{12}{2^N \sqrt{\pi}} \B{r}^{\rm T} \cdot {\rm E} [\left\| \B{x} \right\|_2^2 \cdot \B{x}]  \\
&= \frac{1}{2^N} \frac{-4}{\sqrt{\pi}} (\B{r}^{\rm T} {\rm E} [\B{x}\circ \B{x} \circ \B{x}] )  + \frac{12}{2^N \sqrt{\pi}} \B{r}^{\rm T} \cdot {\rm E} [\left\| \B{x} \right\|_2^2 \cdot \B{x}]  + \frac{48}{2^N} \frac{1}{\pi^{\frac{3}{2}}} \sum_{ \scriptscriptstyle  j \neq i, i \neq l, j \neq l 
} {\rm E}\left[r_i  x_i r_j  x_j  r_l  x_l \right]  \\
&~~~~~    -\frac{12}{2^N \sqrt{\pi}} \B{r}^{\rm T} \cdot {\rm E} [\left\| \B{x} \right\|_2^2 \cdot \B{x}]    \\
& =  \frac{1}{2^N} \frac{-4}{\sqrt{\pi}} (\B{r}^{\rm T} {\rm E} [\B{x}\circ \B{x} \circ \B{x}] )   +  \frac{48}{2^N} \frac{1}{\pi^{\frac{3}{2}}} \sum_{ \scriptscriptstyle  j \neq i, i \neq l, j \neq l 
}  r_i r_j r_l {\rm E} \left[   x_i   x_j    x_l \right], 
\end{aligned}
\end{equation}
which follows from  
\begin{equation}
\begin{aligned}
&{\rm E}_{\B{x}} \int_{\mathbb{R}_+^N}\frac{8}{\pi^N}{\rm e}^{-\left\|\B{y}\right\|_2^2}   ((\B{r}\circ\B{y})^{\rm T} \B{x})^3  {\rm d}\B{y}  =   \\
& =  \frac{8}{\pi^N}{\rm E}_{\B{x}}   \int_{\mathbb{R}_+^N} {\rm e}^{-\left\|\B{y}\right\|_2^2}    (\sum_i r_i^3 y_i^3 x_i^3 + 3 \sum_{i,j\neq i} r_i^2 y_i^2 x_i^2 r_j y_j x_j   + 6 \sum_{j\neq i \neq l} r_i y_i x_i r_j y_j x_j  r_l y_l x_l)   {\rm d} \B{y} \\
&= \frac{8}{2^N}{\rm E}_{\B{x}}     \left[\sum_i r_i \frac{1}{\sqrt{\pi}} x_i^3 + 3 \sum_{i,j\neq i}  \frac{1}{2} x_i^2 r_j \frac{1}{\sqrt{\pi}} x_j + 6 \sum_{j\neq i \neq l} r_i \frac{1}{\sqrt{\pi}} x_i r_j \frac{1}{\sqrt{\pi}} x_j  r_l \frac{1}{\sqrt{\pi}} x_l\right]  \\
&= \frac{8}{2^N}{\rm E}_{\B{x}}  \left[-\frac{1}{2} \sum_i r_i \frac{1}{\sqrt{\pi}} x_i^3 + \frac{3}{2} \sum_{i,j}  x_i^2 r_j \frac{1}{\sqrt{\pi}} x_j + 6 \sum_{j\neq i \neq l} r_i \frac{1}{\sqrt{\pi}} x_i r_j \frac{1}{\sqrt{\pi}} x_j  r_l \frac{1}{\sqrt{\pi}} x_l\right]  \\
&= -\frac{4}{2^N}{\rm E}_{\B{x}} (\B{r}^{\rm T} {\rm E} [\B{x}\circ \B{x} \circ \B{x}] )   + \frac{12}{2^N \sqrt{\pi}} \B{r}^{\rm T} \cdot {\rm E} [\left\| \B{x} \right\|_2^2 \cdot \B{x}]  +    \frac{48}{2^N} \frac{1}{\pi^{\frac{3}{2}}} \sum_{ \scriptscriptstyle  j \neq i, i \neq l, j \neq l 
} {\rm E}\left[r_i  x_i r_j  x_j  r_l  x_l \right].
\end{aligned}
\end{equation}

Now, after the derivation of the derivatives $P_0'(\B{r})$, $P_0''(\B{r})$ and  $P_0^{(3)}(\B{r})$, we have to perform a summation with respect to $\B{r}\in \{-1,+1\}^N$ of the terms given in (\ref{entropy_order_2}). For this,  we make use of the following  properties:
\begin{align}
\frac{1}{2^N} \sum_{\B{r} \in \{-1,+1\}^N}  \B{r}^{\rm T} \B{A} \B{r} ={\rm tr}(\B{A}),
\end{align}
and
\begin{align}
\frac{1}{2^N} \sum_{\B{r} \in \{-1,+1\}^N} \B{r}^{\rm T} \B{A} \B{r} \cdot \B{r}^{\rm T} \B{B} \B{r} = {\rm tr}( \B{A} \cdot {\rm nondiag}(\B{B} + \B{B}^{\rm T})) + {\rm tr} (\B{A}) {\rm tr} (\B{B}),
\end{align}
which can be verified by expanding the left-hand side and identifying the non-zero terms.
We start by the second order expression of (\ref{entropy_order_2})
\begin{equation}
\begin{aligned}
\sum_{\B{r} \in \{-1,+1\}^N}  \frac{P_0'(\B{r})^2}{2P_0(\B{r})} &=  \sum_{\B{r} \in \{-1,+1\}^N} \frac{2}{2^N \cdot \pi} (\B{r}^{\rm T} {\rm E}[\B{x}]^{\rm T})^2 \\
          &= \sum_{\B{r} \in \{-1,+1\}^N} \frac{2}{2^N \cdot \pi } \B{r}^{\rm T} \underbrace{{\rm E}[\B{x}] {\rm E}[\B{x}]^{\rm T}}_{\B{A}}\B{r}  \\
          &= \frac{2}{\pi} \left\| \B{x} \right\|_2^2.       
\end{aligned}
\label{sum_der1}
\end{equation}
Then, we consider the different terms of the fourth-order derivatives, starting with 
\begin{equation}
\begin{aligned}
\sum_{\B{r} \in \{-1,+1\}^N}  \frac{P_0'(\B{r})^4}{12P_0(\B{r})^3 } &=  \sum_{\B{r} \in \{-1,+1\}^N} \frac{4}{2^N \cdot 3\pi^2} (\B{r}^{\rm T} {\rm E}[\B{x}]^{\rm T})^4 \\
          &= \sum_{\B{r} \in \{-1,+1\}^N} \frac{4}{2^N \cdot 3\pi^2}  \B{r}^{\rm T} \underbrace{{\rm E}[\B{x}] {\rm E}[\B{x}]^{\rm T}}_{\B{A}}\B{r} \B{r}^{\rm T} \underbrace{{\rm E}[\B{x}] {\rm E}[\B{x}]^{\rm T}}_{\B{B}}\B{r}   \\
          &= \frac{4}{\pi^2} {\rm tr} ({\rm nondiag} ({\rm E}[\B{x}]{\rm E}[\B{x}]^{\rm T})^2)   + \frac{4}{3\pi^2} \left\|{\rm E}[\B{x}]\right\|_4^4. 
\end{aligned}
\label{sum_der2}
\end{equation}
On the other hand, we have
\begin{equation}
\begin{aligned}
\sum_{\B{r} \in \{-1,+1\}^N}  \frac{P_0''(\B{r})^2}{8 P_0(\B{r}) } &=  \sum_{\B{r} \in \{-1,+1\}^N} \frac{2}{2^N \cdot \pi^2}  \B{r}^{\rm T} {\rm nondiag} \left({\rm E} [\B{x}\B{x}^{\rm T}]\right) \B{r} \B{r}^{\rm T} {\rm nondiag} \left({\rm E} [\B{x}\B{x}^{\rm T}]\right) \B{r} \\
          &= \frac{4}{\pi^2}  {\rm tr} ({\rm nondiag} ({\rm E}[\B{x}\B{x}^{\rm T}])^2).  
\end{aligned}
\label{sum_der3}
\end{equation}
Next, we get 
\begin{equation}
\begin{aligned}
\sum_{\B{r} \in \{-1,+1\}^N}  \frac{P_0'(\B{r})^2P_0''(\B{r})}{4P_0(\B{r})^2}  &=  \sum_{\B{r} \in \{-1,+1\}^N} \frac{4}{2^N \cdot \pi^2}  \B{r}^{\rm T} {\rm nondiag} \left({\rm E} [\B{x}\B{x}^{\rm T}]\right) \B{r} \B{r}^{\rm T} {\rm nondiag} \left({\rm E} [\B{x}]{\rm E}[\B{x}]^{\rm T}\right) \B{r} \\
          &= \frac{8}{\pi^2} {\rm tr} ({\rm nondiag} ({\rm E}[\B{x}]{\rm E}[\B{x}]^{\rm T}) {\rm nondiag} ({\rm E}[\B{x}\B{x}^{\rm T}])). 
\end{aligned}
\label{sum_der4}
\end{equation}
Further, we obtain
\begin{equation}
\begin{aligned}
& \sum_{\B{r} \in \{-1,+1\}^N} \frac{P_0'(\B{r})P_0^{(3)}(\B{r})}{6P_0(\B{r})}  =  \\  
& \frac{1}{6\cdot 2^N} \sum_{\B{r} \in \{-1,+1\}^N}   \left( \frac{2}{\sqrt{\pi}}  \B{r}^{\rm T} {\rm E} [\B{x}]  \right)\left( \frac{-4}{\sqrt{\pi}} (\B{r}^{\rm T} {\rm E} [\B{x}\circ \B{x} \circ \B{x}] )   +  \frac{48}{2^N} \frac{1}{\pi^{\frac{3}{2}}} \sum_{\scriptscriptstyle  j \neq i, i \neq l, j \neq l}  r_i r_j r_l {\rm E} \left[  x_i  x_j  x_l  \right]  \right)   \\
&=  - \frac{4}{3 \pi} {\rm E}[\B{x}]^{\rm T} \cdot {\rm E}[\B{x} \circ \B{x}  \circ \B{x}]   + \frac{16}{\pi^2} \sum_{\B{r}\in\{\pm 1 \}^N} \sum_{\scriptscriptstyle  j \neq i, i \neq l, j \neq l, k}   r_i r_j r_l r_k {\rm E}[x_k] {\rm E} \left[ x_i  x_j  x_l  \right]  \\    
&=  - \frac{4}{3 \pi} {\rm E}[\B{x}]^{\rm T} \cdot {\rm E}[\B{x} \circ \B{x}  \circ \B{x}]+0.       
\end{aligned}
\label{sum_der5}
\end{equation}

We plug the results from (\ref{sum_der1})--(\ref{sum_der5}) into the   expression for the entropy in (\ref{entropy_order_2}), to get finally
\begin{equation}
\begin{aligned}
H({\rm sign}(\varepsilon \B{x}+ \B{\eta}))  = & 2N\ln 2 -\frac{2}{\pi}\varepsilon^2 \left\|{\rm E}[\B{x}]\right\|_2^2-\varepsilon^4\left(  \frac{4}{\pi^2} {\rm tr}\left(\left({\rm nondiag}(\B{C}_{\B{x}}) \right)^2\right) \right.  \\ 
& \left. -\frac{4}{3\pi}\left\|{\rm E}[\B{x}]\circ {\rm E}[\B{x}\circ\B{x}\circ \B{x}] \right\|_1^1+\frac{4}{3\pi^2}\left\|{\rm E}[g(\B{x})]\right\|_4^4\right)  + \Delta H({\rm sign}(\varepsilon \B{x}+ \B{\eta})).
\end{aligned}
\end{equation}

Now, for the complex valued case, we easily obtain a similar result, since any $N$-dimensional vector can be represented as a real-valued $2N$-dimensional vector, where we split the vector into the real part and the imaginary part and write them as: 

\begin{equation}
\begin{aligned}
\B{x}^r= \left[
\begin{array}{c}
	{\rm Re } \{ x \}  \\
	{\rm Im } \{x \} 
\end{array} \right].
\end{aligned}
\end{equation} 
 This implies that the vector norms can be defined as

\begin{equation}
\begin{aligned}
\left\| \B{x} \right\|_4^4 = \sum_k {\rm Re}\{ x_k \}^4 + {\rm Im}\{ x_k \}^4. 
\end{aligned}
\end{equation}  
On the other hand, the covariance matrix can be written as
\begin{equation}
\begin{aligned} 
\B{C}_{\B{x}} = {\rm E}[\B{x} \B{x}^{\rm H}] - {\rm E} [\B{x}]{\rm E}[\B{x}^{\rm H}].
\end{aligned} 
\end{equation}
Under the assumption that $\B{x}$ is circularly distributed, i.e., ${\rm E}[\B{x}\B{x}^{\rm T}]={\rm E}[\B{x}][\B{x}^{\rm T}]$, then we have
\begin{align} 
\B{C}_{\B{x}^r} = {\rm E} [\B{x}^{r} \B{x}^{r,{\rm T}}] =  \frac{1}{2} \left[   
\begin{array}{cc}
{\rm Re} \{ \B{C}_{\B{x}} \} 	&  {\rm Im} \{ \B{C}_{\B{x}} \} \\
-{\rm Im} \{ \B{C}_{\B{x}} \} 	&  {\rm Re} \{ \B{C}_{\B{x}} \} 
\end{array}
\right],   
\end{align} 
where we note that the matrix ${\rm Im} \{ \B{C}_{\B{x}} \} $ is a skew-symmetric matrix,  i.e., ${\rm Im} \{ \B{C}_{\B{x}}^{\rm T} \}=-{\rm Im} \{ \B{C}_{\B{x}} \}$ and has diagonal elements equal to zero, thus we have
\begin{equation}
\begin{aligned}
 ({\rm nondiag}(\B{C}_{\B{x}^{r}}))^2  &  =  \frac{1}{2} ({\rm nondiag} ({\rm Re} \{ \B{C}_{\B{x}}\}))^2 - \frac{1}{2}({\rm Im} \{ \B{C}_{\B{x}} \} )^2      \\
                                       &  =  \frac{1}{2} ({\rm nondiag} ({\rm Re} \{ \B{C}_{\B{x}}\}))^2 - \frac{1}{2} ({\rm nondiag} ({\rm Im} \{ \B{C}_{\B{x}} \}))^2  \\
                                       &  =  \frac{1}{2} ({\rm nondiag} (\B{C}_{\B{x}}) )^2.
\end{aligned}
\end{equation}
 Therefore, we can conclude that the formula remains the same except that we replace $({\rm nondiag}({\rm E}[\B{x} \B{x}^{\rm H}]))^2$ by \\
  $\frac{1}{2}({\rm nondiag}({\rm E}[\B{x} \B{x}^{\rm H}]))^2$, and the norms retain their same definition.
Now, we turn to the question of the necessary conditions such that the second order approximation is valid.
The condition $\textrm{E}_{\B{x}} [\left\|\B{x}\right\|_4^{4+\alpha}]<\gamma$ for some finite constants $\alpha,\gamma>0$  stated by the theorem is necessary, so that the remainder term of the expansion given by
\begin{equation}
\Delta H({\rm sign}(\varepsilon \B{x}+ \B{\eta})) =\textrm{E}_{\B{x}}[o(\left\|\B{x}\right\|_4^4 \varepsilon^4 )]
\end{equation}
satisfies
\begin{equation} 
 \lim_{ \varepsilon^4  \rightarrow 0}\frac{\Delta H(\B{x},\B{y})}  {\varepsilon^4} =0,
 \end{equation} 
 since
\begin{eqnarray}
\begin{aligned}
\!\Delta H(\B{x},\B{r})&=	\textrm{E}_{\B{x}}[o(\left\|\B{x}\right\|_4^4 \varepsilon^4  )] \nonumber  \\
	&\leq  \textrm{E}_{\B{x}}[(\left\|\B{x}\right\|_4^4 \varepsilon^4 )^{1+\frac{\alpha'}{4}}], \textrm{ for some } \alpha' \in ]0,\alpha] \nonumber  \\
	&\leq \textrm{E}_{\B{x}}[\left\|\B{x}\right\|_4^{4+\alpha'}] \varepsilon^{4+\alpha'} \nonumber  \\
		&\leq \textrm{E}_{\B{x}}[\left\|\B{x}\right\|_4^{4+\alpha}]^{\frac{4+\alpha'}{4+\alpha}}  \varepsilon^{4+\alpha'}    \textrm{(H\"older's inequality)} \nonumber\\
			&\leq \gamma^{\frac{4+\alpha'}{4+\alpha}} \varepsilon^{4+\alpha'} \nonumber  \\
		&= o(\varepsilon^4).
	\end{aligned}
\end{eqnarray}

 This yields  the result stated by the theorem and completes the proof. 

 \bibliographystyle{IEEEbib}
\bibliography{IEEEabrv,references}
\end{document}